\documentclass[12pt]{article}
\pdfoutput=1
\usepackage{jheppub}
\usepackage{mathtools,amsmath,amssymb,amsthm,physics,bm,float}
\usepackage[utf8]{inputenc}
\usepackage{graphicx}
\usepackage{enumitem}
\usepackage{hyperref}

\begin{document}

\DeclarePairedDelimiter{\ceil}{\lceil}{\rceil}
\newtheorem{theorem}{Theorem}
\newtheorem{lemma}[theorem]{Lemma}
\newtheorem{corollary}[theorem]{Corollary}
\newcommand{\be}{\begin{equation}}
\newcommand{\ee}{\end{equation}}
\newcommand{\bi}{\begin{itemize}}
\newcommand{\ei}{\end{itemize}}
\newcommand{\mathbbm}[1]{\text{\usefont{U}{bbm}{m}{n}#1}}
\def\ba#1\ea{\begin{align}#1\end{align}}
\def\bg#1\eg{\begin{gather}#1\end{gather}}
\def\bm#1\em{\begin{multline}#1\end{multline}}
\def\bmd#1\emd{\begin{multlined}#1\end{multlined}}
\setlength{\intextsep}{-1ex} 

\def\XD#1{{\color{magenta}{ [#1]}}}
\def\SAM#1{{\color{red}{ [#1]}}}
\def\WW#1{{\color{blue}{ [#1]}}}

\def\a{\alpha}
\def\b{\beta}
\def\c{\chi}
\def\C{\Chi}
\def\d{\delta}
\def\D{\Delta}
\def\e{\epsilon}
\def\ve{\varepsilon}
\def\g{\gamma}
\def\G{\Gamma}
\def\h{\eta}
\def\k{\kappa}
\def\l{\lambda}
\def\L{\Lambda}
\def\m{\mu}
\def\n{\nu}
\def\p{\phi}
\def\P{\Phi}
\def\vp{\varphi}
\def\q{\theta}
\def\Q{\Theta}
\def\r{\rho}
\def\s{\sigma}
\def\S{\Sigma}
\def\t{\tau}
\def\u{\upsilon}
\def\U{\Upsilon}
\def\w{\omega}
\def\W{\Omega}
\def\x{\xi}
\def\X{\Xi}
\def\y{\psi}
\def\Y{\Psi}
\def\z{\zeta}

\newcommand{\la}{\label}
\newcommand{\ci}{\cite}
\newcommand{\re}{\ref}
\newcommand{\er}{\eqref}
\newcommand{\se}{\section}
\newcommand{\sse}{\subsection}
\newcommand{\ssse}{\subsubsection}
\newcommand{\fr}{\frac}
\newcommand{\na}{\nabla}
\newcommand{\pa}{\partial}
\newcommand{\td}{\tilde}
\newcommand{\wtd}{\widetilde}
\newcommand{\ph}{\phantom}
\newcommand{\eq}{\equiv}
\newcommand{\wg}{\wedge}
\newcommand{\cd}{\cdots}
\newcommand{\nn}{\nonumber}
\newcommand{\qu}{\quad}
\newcommand{\qqu}{\qquad}
\newcommand{\lt}{\left}
\newcommand{\rt}{\right}
\newcommand{\lra}{\leftrightarrow}
\newcommand{\ol}{\overline}
\newcommand{\ap}{\approx}
\renewcommand{\(}{\left(}
\renewcommand{\)}{\right)}
\renewcommand{\[}{\left[}
\renewcommand{\]}{\right]}
\newcommand{\<}{\langle}
\renewcommand{\>}{\rangle}
\newcommand{\Hc}{\mathcal{H}_{code}}
\newcommand{\HR}{\mathcal{H}_R}
\newcommand{\HRb}{\mathcal{H}_{\ol{R}}}
\newcommand{\lan}{\langle}
\newcommand{\ran}{\rangle}
\newcommand{\Hra}{\mathcal{H}_{W_\alpha}}
\newcommand{\Hrba}{\mathcal{H}_{\ol{W}_\alpha}}

\newcommand{\bH}{{\mathbb H}}
\newcommand{\bR}{{\mathbb R}}
\newcommand{\bZ}{{\mathbb Z}}
\newcommand{\cA}{{\mathcal A}}
\newcommand{\cB}{{\mathcal B}}
\newcommand{\cC}{{\mathcal C}}
\newcommand{\cE}{{\mathcal E}}
\newcommand{\cI}{{\mathcal I}}
\newcommand{\cN}{{\mathcal N}}
\newcommand{\cO}{{\mathcal O}}
\newcommand{\zb}{{\bar z}}

\newcommand{\Area}{\operatorname{Area}}
\newcommand{\ext}{\operatorname*{ext}}
\newcommand{\total}{\text{total}}
\newcommand{\bulk}{\text{bulk}}
\newcommand{\brane}{\text{brane}}
\newcommand{\matter}{\text{matter}}
\newcommand{\Wald}{\text{Wald}}
\newcommand{\anomaly}{\text{anomaly}}
\newcommand{\extrinsic}{\text{extrinsic}}
\newcommand{\gen}{\text{gen}}
\newcommand{\mc}{\text{mc}}

\newcommand{\T}[3]{{#1^{#2}_{\ph{#2}#3}}}
\newcommand{\Tu}[3]{{#1_{#2}^{\ph{#2}#3}}}
\newcommand{\Tud}[4]{{#1^{\ph{#2}#3}_{#2\ph{#3}#4}}}
\newcommand{\Tdu}[4]{{#1_{\ph{#2}#3}^{#2\ph{#3}#4}}}
\newcommand{\Tdud}[5]{{#1_{#2\ph{#3}#4}^{\ph{#2}#3\ph{#4}#5}}}
\newcommand{\Tudu}[5]{{#1^{#2\ph{#3}#4}_{\ph{#2}#3\ph{#4}#5}}}

\newcommand{\bs}{\boldsymbol}
\newcommand{\bfRho}{\bs{\rho}}
\newcommand{\bfS}{\textbf{S}}

\title{
Entanglement Negativity Transitions in Chaotic Eigenstates
}
\author{Sean McBride and}
\author{Fernando Iniguez}
\affiliation{Department of Physics, University of California, Santa Barbara, CA 93106, USA}
\emailAdd{seanmcbride@ucsb.edu}
\emailAdd{finiguez@ucsb.edu}

\abstract{It was recently noted that the entanglement entropy for a subsystem of a chaotic eigenstate exhibits an enhanced correction when the subsystem approaches a phase transition at half the total system size. This enhanced correction was derived for general subsystems by Dong and Wang by summing over noncrossing permutations, which can be thought of as ``saddles'' either in a sum emerging from averaging over Wick contractions or in an analogous gravitational calculation. We extend these results to the case of entanglement negativity, an entanglement measure defined on a bipartite density matrix. We focus on a particular transition previously studied in a toy model of JT gravity, one for which the sum over permutations was found to give similar (or even stronger) enhanced corrections. We derive and resum the relevant permutations to give a form for the averaged negativity spectrum, reproducing the gravitational answer for some quantities and finding tension with other quantities, namely the partially transposed entropy. Along the way, we extend the results of Dong and Wang to the case of $n < 1$ R\'enyi entropy, showing that it always receives volume law corrections.
}
\maketitle

\section{Introduction}

The application of ideas from quantum chaos to gravitational settings has been particularly fruitful. Gravitational observables have been shown to be well approximated by observables obeying the eigenstate thermalization hypothesis (ETH) \cite{PhysRevA.43.2046, PhysRevE.50.888, Srednicki:1995pt, Srednicki_1999}. This is due to the fact that a holographic quantum field theory with a semiclassical Einstein gravity dual is expected to be maximally chaotic, i.e. it saturates the bound of \cite{Maldacena:2015waa}, up to higher derivative/stringy corrections which take one away from this regime. The power of ETH is that it allows us to approximate observables in the microcanonical ensemble by an observable's long-time quantum expectation value. The resulting microcanonical expectation value should resemble that of the canonical ensemble, up to corrections expected to be suppressed in the system size $V$ by the thermodynamic ensembles’ equivalence at large $N$. This gives a quantitative idea of the process of thermalization in isolated quantum many-body systems.

From the perspective of subsystem ETH \cite{Garrison:2015lva, Dymarsky:2016ntg, Dymarsky:2017zoc, Lashkari:2017hwq, Lu:2017tbo}, for a subsystem with volume fraction $f < 1/2$, the corrections to ETH are suppressed in system size. Formally, this means the trace-norm distance between the canonical and microcanonical density matrices vanishes in the large volume/thermodynamic limit, implying that off-diagonal matrix elements of operators vanishes and expectation values are roughly thermal. This line of thinking is expected to apply to R\'enyi entropies. 

When $f = 1/2$ exactly, the usual wisdom would say there's an $\mathcal{O}(1)$ correction to the R\'enyi entropies. One way of seeing this is that there exists a phase transition in the R\'enyi entropy at $f = 1/2$, and the correction at this phase transition should be given by the uncertainty in choosing between an $\mathcal{O}(1)$ number of equivalent dominant phases. However, in a model-specific result, a correction to the entanglement entropy of $\mathcal{O}(\sqrt{V})$ was observed in \cite{VidmarRigol}, a correction derived in \cite{Murthy:2019qvb} and explicated in \cite{Dong:2020iod}. In particular, the von Neumann entropy of a subregion $A$ with volume fraction $f = 1/2$ takes the form
\be
S_A = \frac{S(E)}{2} - \sqrt{\frac{C_V}{2 \pi}} + \mathcal{O}(a)
\label{eq:vNcorrection}
\ee
where $S(E)$ is the thermodynamic entropy at energy $E$ and $C_V$ is the heat capacity at constant volume. As $C_V$ is extensive in the system size, the correction is ``enhanced'' to $\mathcal{O}(\sqrt{V})$. This formula is valid in the large volume limit, where $\sqrt{V} \gg a$, $a$ being the area of the splitting surface. 

In a parallel story, the attempt to match results from tensor networks with the gravitational path integral led to the understanding of ``fixed area states'' \cite{Dong:2018seb, Akers:2018fow}. These states, which are eigenstates of the area operator in semiclassical gravity, have a flat entanglement spectrum, up to fluctuations about a fixed saddle point which can na\"ively be at most $\mathcal{O}(1)$ in units of $G_N$, where $G_N \ll 1$. One can think of these fluctuations as the difference in the ``canonical'' ensemble where one fixes the canonical conjugate to the area operator, namely the relative boost between the entanglement wedges of the two sides \cite{Dong:2022ilf}, and a ``microcanonical'' ensemble where the eigenvalue of the area operator is fixed at its most probable value.

It was noted in \cite{Marolf:2020vsi, Dong:2020iod} that the universal enhanced correction to the entanglement entropy also appears in fixed area states near transition, where the ``transition'' in this context occurs due to a competition between two extremal surfaces. One way of understanding this correction is that near transition we no longer care about fluctuations about a fixed saddle, but instead we care about resumming an infinite number of saddles which appear in the sum over topologies in the replicated geometries. Both of these results match with a more detailed calculation of the same quantity in \cite{Penington:2019kki}, where in a model of Jackiw-Teitelboim (JT) gravity + end-of-the-world (EOW) branes a particular subsystem entropy $S(\rho_R)$ had the form
\be
S(\rho_R) = \log k - \sqrt{\frac{2 \pi}{\beta}} + \mathcal{O} ( \log  \beta ).
\ee
This $\sqrt{1/\beta}$ correction is analogous to the $\sqrt{C_V}$ correction in chaotic eigenstates. Here we've set Newton's constant (which is analagous to $N$) to one, but it can be restored via $\beta \rightarrow G_N \beta$.

Recently, it was shown by \cite{Dong:2021oad} that similar enhanced corrections exist near transitions in entanglement negativity, a tripartite entanglement measure defined on a bipartite density matrix $\rho_{A_1 A_2}$. In particular, the logarithmic negativity was shown to have the following form at transition:
\be
\mathcal{E}(\rho_{R_1 R_2}) = \log k_2 - \frac{\pi^2}{8 \beta} + \mathcal{O}(\log \beta),
\ee
for two subsystems $R_1$ and $R_2$ with $R_1 \cup R_2 = R$. Further corrections were derived for measures descending from a R\'enyi version of negativity.

There exists a rich phase diagram for entanglement negativity in holographic states, and we show that a similar phase diagram exists for a generic chaotic eigenstate. Our aim is to systematically derive the corrections at transitions in this phase space. There are two possible transitions, but as was explored in \cite{Dong:2021oad} we only expect interesting behavior near one of the transitions, for reasons we'll recapitulate in the main text. 

The outline of the paper is as follows. In section \ref{sec:preface} we review the derivation of \eqref{eq:vNcorrection}, in particular the resolvent formalism of \cite{Dong:2020iod}. In section \ref{sec:neg} we review the various negativity measures discussed in \cite{Dong:2021oad} and the sum over relevant permutations for the phase transition of interest. In section \ref{sec:trans} we compute corrections to the entanglement measures of interest. We conclude with some discussion and future directions.

\section{Diagrammatics for Chaotic Subsystems}
\label{sec:preface}

We first review the formalism of \cite{Murthy:2019qvb, Dong:2020iod}, which was used to compute the universal form of corrections to the entanglement entropy of a subsystem at transition. We focus on \cite{Dong:2020iod}, as their formalism more easily generalizes to our future calculations. Readers familiar with their formalism may skip this section, whose only purpose is to make this work self-contained.

A generic eigenstate $\ket{E}$ of a Hamiltonian $H$ defined on a bipartite system such that $\mathcal{H} = \mathcal{H}_A \otimes \mathcal{H}_B$ can be Schmidt decomposed via
\be
\ket{E} = \sum_{iJ} M_{iJ} \ket{E_i}_A \otimes \ket{E_J}_B,
\ee
where $\ket{E_i}$ and $\ket{E_J}$ denote eigenstates of the subsystem Hamiltonians $H_A$ and $H_B$, respectively. As is convention, we use lowercase indices for states of $A$ and uppercase indices for states of $B$. ETH instructs us to think of $M_{iJ}$ as a Gaussian random variable with zero mean and energy banded with width $\Delta$ \cite{Deutsch2, LuGrover}. In particular, for a system with spatial dimension $d \geq 2$, we have the ansatz
\be
M_{iJ} = e^{-S(E_{Ai}+E_{BJ})/2} \left( \frac{e^{-\epsilon^2/2\Delta^2}}{\sqrt{2\pi} \Delta} \right)^{1/2} C_{iJ},
\ee
where $\epsilon = E_{i} + E_{J} - E$ is the deviation from the total microcanonical energy. When averaged over a small energy band in $E_A$ and $E_B$, the random coefficients $C_{iJ}$ satisfy
\be
\overline{C_{iJ}} = 0, \quad \overline{C_{iJ}C_{i'J'}} = \delta_{ii'} \delta_{JJ'}.
\ee
The effects of finite $\Delta$ will not affect the current and future computation, so we work in the limit $\Delta \rightarrow 0$, where we approximate
\be
M_{iJ} \approx e^{-S(E)/2} C_{iJ}.
\ee
This approximation assumes the true density of states in a narrow energy band is well approximated by the thermodynamic entropy in the canonical ensemble. To leading order in the system volume, we can further approximate the density of states of the total system as the product of the density of states of the subsystems $A$ and $B$, evaluated at the subsystem energy $E_A$. In other words,
\be
S(E) \approx S_A(E_A) + S_B(E - E_A).
\ee
This leads to the following form for a subsystem density matrix $\rho_A$
\be
\rho_A = \frac{1}{\mathcal{N}}\sum_{E_i - 2\Delta < E_j < E_i + 2\Delta } \sum_{E - E_i - \Delta < E_J < E - E_i + \Delta} C_{iJ} C_{jJ} \ket{E_i}_A \bra{E_j}_A.
\label{eq:unaveragedrho}
\ee
The double sum takes into account energies in a region of width $2\Delta$. Averaging over the $C_{ij}$'s gives the averaged subsystem density matrix 
\be
\overline{\rho_A} = \frac{1}{\mathcal{N}} \sum_i d_B(E-E_i) \ket{E_i}_A \bra{E_i}_A,
\ee
where the normalization is given by
\be
\mathcal{N} = \sum_i d_A(E_i) d_B(E-E_i).
\ee
Here $d_A$ and $d_B$ are the degeneracies at a given energy. As a shorthand and as a motivation for our future computation, we can instead write
\be
d_A(E_i) = e^{S_A(E_i)} \equiv e^{S_A}; \quad d_B(E - E_i) = e^{S_B(E-E_i)} \equiv e^{S_B}.
\ee
As our goal is to compute subsystem von Neumann entropy, we should proceed by generalizing this procedure to compute $\Tr \overline{\rho_A^n}$. Before averaging, from \eqref{eq:unaveragedrho} we have
\be
\rho_A^n = \frac{1}{\mathcal{N}^n} \sum_{E_{i_1}}\sum_{i_2, \cdots, i_{n+1}; J_1, \cdots, J_n} \prod_{m=1}^{n} C_{i_m J_m} C_{i_{m+1}J_m} \ket{E_{i_1}}_A \bra{E_{i_{n+1}}}_A,
\ee
where the second sum is understood to be over a strip of width $2n \Delta$, but we assume $\Delta$ vanishes quickly enough at finite $n$ that this isn't a significant effect.

The difference between $\overline{\log \Tr \left( \rho_A \right)^n}$ and $\log \Tr \overline{ \left( \rho_A \right)^n}$ is exponentially suppressed in the system volume~\cite{Lu:2017tbo}, so our goal will be to compute $\Tr \overline{ \left( \rho_A \right)^n}$, as it is a more tractable calculation. This involves a sum over Wick contractions, as we assume higher point connected correlations of the $C_{iJ}$'s vanish. The result is 
\be
\Tr \overline{(\rho_A)^n} = \begin{cases}
\frac{1}{\mathcal{N}^n} e^{S_A + nS_B}{}_2F_1 \left( 1-n, -n; 2; e^{S_A-S_B} \right), S_A < S_B \\
\frac{1}{\mathcal{N}^n} e^{nS_A + S_B} {}_2F_1 \left( 1-n, -n; 2; e^{S_B - S_A} \right), S_A > S_B.
\end{cases}
\ee
As a sanity check, we recover $S_n(\rho_A) = S/2 + \mathcal{O}(1)$ where $S_A = S_B = S/2$ at $f = 1/2$. The derivation of this expression from the resolvent sum over noncrossing permutations is given in Appendix \ref{app:resolvent}.

To study the corrections to this quantity at transition, we upgrade the putative constant density of states to an integral over an energy dependent density of states. In other words, we send
\be
S_A \rightarrow S_A(E_A), \quad S_B \rightarrow S_B(E - E_A)
\ee
and integrate over $E_A$. The new averaged trace is given by
\be
\Tr \overline{ (\rho_A)^n} = \frac{1}{\mathcal{N}^n} \int dE_A e^{S_A(E_A) + S_B(E - E_A)} G_n(E_A),
\label{eq:tr1}
\ee
where $G_n(f,E_A)$ encompasses the $n$-dependent piece of the trace:
\be
G_n(E_A) = \begin{cases} e^{ (n-1)S_B(E-E_A)}{}_2F_1 \left( 1-n, -n; 2; e^{S_A(E_A)-S_B(E-E_A)} \right), S_A(E_A) < S_B(E-E_A) \\
e^{(n-1)S_A(E_A)} {}_2F_1 \left( 1-n, -n; 2; e^{S_B(E-E_A) - S_A(E_A)} \right), S_A(E_A) > S_B(E-E_A) \end{cases}
\label{eq:GnRenyi}
\ee
and the normalization is now 
\be
\mathcal{N} = \int dE_A e^{S_A(E_A) + S_B(E - E_A)}.
\ee
From this, we can directly calculate the ensemble averaged R\'enyi entropies $S_n(\rho_A)$:
\be
\overline{S_n} = \frac{1}{1-n} \log \left( \frac{1}{\mathcal{N}^n} \int  dE_A e^{S_A(E_A) + S_B(E - E_A)} G_n(E_A)  \right).
\ee

\subsection{Saddle Point Analysis}

We make the ansatz that the entropy is extensive in the subsystem size, that is
\be
S_A(E_A) = f V s \left( \frac{E_A}{fV}\right), \quad S_B(E - E_A) = (1-f) V s \left( \frac{E-E_A}{(1-f)V}\right),
\ee
where $f \equiv V_A/V$ is the volume fraction, $s(e)$ is the entropy density as a function of the energy density $e$, and the other factors come from dimensional analysis. We're mainly interested in what happens at the transition $f = 1/2$. The ``featureless'' or infinite temperature case is when $s(e) = 1$ such that all subsystem entropies are proportional to subsystem volume. We're only interested in the corrections from finite temperature, which can be thought of as the difference between the answer in the canonical ensemble and the microcanonical ensemble. The microcanonical R\'enyi entropy is the contribution of the global ``unaveraged'' microcanonical state $\rho = \sum_{E - \Delta < E_i < E + \Delta} \ket{E_i} \bra{E_i}:$
\be
S^{MC}_n = \frac{1}{1-n} \log \left( \frac{1}{\mathcal{N}^n} \int  dE_A e^{S_A(E_A) + n S_B(E - E_A)}  \right).
\ee
We are interested in the correction away from the dominant microcanonical saddle, so we are interested in computing the following quantity:
\be
\overline{S_n} - S^{MC}_n = \frac{1}{1-n} \ln \left( \frac{\int dE_A \exp (F_1(E_A))}{\int dE_A \exp (F_2(E_A))}\right),
\ee
where $F_1(E_A)$ and $F_2(E_A)$ are functions defined by
\ba
F_1(E_A) &=  f V s \left( \frac{E_A}{fV}\right) + (1-f) V s \left( \frac{E-E_A}{(1-f)V}\right) + \ln G_n(E_A) \nonumber \\
F_2(E_A) &= f V s \left( \frac{E_A}{fV}\right) + n(1-f) V s \left( \frac{E-E_A}{(1-f)V}\right).
\ea
As both functions scale with volume, we can perform a saddle point analysis. The saddle point equations for these functions are
\ba
s' \left( \frac{E_1}{fV}\right) &= s' \left( \frac{E - E_1}{(1-f)V}\right) - \frac{G'_n(f,E_1)}{G_n(f,E_1)} \nonumber \\
s'\left(\frac{E_2}{fV} \right) &= ns' \left( \frac{E - E_2}{(1-f)V}\right),
\label{eq:renyisaddles}
\ea
where $E_1$ and $E_2$ are the saddle point energies of $F_1(E_A)$ and $F_2(E_A)$, respectively. The analysis of these saddle point equations was done in totality for $n > 1$ in \cite{Dong:2020iod}.\footnote{See also \cite{Kudler-Flam:2021alo} for a similar study of relative entropy with the same ansatz.} Here we fill in a small gap and study the case of $n < 1$. This will be useful later when we are computing analytic continuations of R\'enyi negativities below $n = 1$.

\subsection{Corrections at Transition for $n < 1$}

$s(x)$ is a monotonically increasing function of $x$ with a monotonically decreasing first derivative (take $s(x) = \sqrt{x}$ as a concrete example). For the case $n < 1$, we can therefore write the iequality
\be
\frac{E_2}{f} > \frac{E - E_2}{1-f},
\ee
which immediately implies
\be
S_A(E_2) > \frac{f}{1-f} S_B(E-E_2),
\ee
and therefore $S_A(E_1) > S_B(E- E_1)$ for $f > 1/2$. 

The first thing to notice is that there is only one saddle point for both $F_1(E_A)$ and $F_2(E_A)$, as $G_n(f,E_A)$ is now a strictly concave function. The single saddle for $F_1(E_A)$ depends sensitively on the saddle point of $G_n(f,E_A)$, which itself only depends on the crossover point between the two hypergeometrics. As the crossover point is completely determined by the $n$-independent quantity
\be
S_A(E_A) - S_B(E-E_A)
\ee
and the rest of the $E_1$ saddle point equation is independent of $n$, the full saddle similarly becomes completely independent of $n$. This should be contrasted with the obviously $n$-dependent saddle point of $F_2(E_A)$. This difference will generically cause the two saddles to differ by an $\mathcal{O}(1)$ factor, so for all volume fractions we expect the different in R\'enyi entropies to be volume law:
\be
\overline{S_n} - S_n^{MC} = \mathcal{O}(V), \quad n < 1.
\ee
Note that this applies for all volume fractions, implying that the $n < 1$ R\'enyi entropies do not obey the principle of canonical typicality.

This clarifies a conceptual point. For $n \rightarrow 1^+$, the $\sqrt{V}$ correction lies in between an exponentially suppressed $\mathcal{O}(e^{-cV})$ region ($f < 1/2$) and a strongly enhanced $\mathcal{O}(V)$ region ($f > 1/2$). Why, then, do we not get a similar enhancement for $n \rightarrow 1^-$? The answer is that the dominant behavior in $F_1(E_A)$, which previously supplied the emergent ``soft mode'' for the flat interval between two saddles, becomes independent of the R\'enyi index. This nonanalyticity might be worrying if one is used to a R\'enyi entropy analytic in $n$, but the thermodynamic limit breaks this assumption. The form of these corrections agrees with the analysis of a gravitational model in Appendix C of \cite{Dong:2021oad}.

\section{Entanglement Negativity}
\label{sec:neg}
In this section we compute similar quantities as \cite{Dong:2020iod} for entanglement negativity measures. We begin by reviewing some salient properties of entanglement negativity and its utility as a tripartite measure of entanglement before diving into the calculation.

\subsection{Review of Negativity}
Entanglement negativity refers to an entanglement measure based on properties of the partial transpose operation applied to a bipartite density matrix $\rho_{A_1 A_2}$, defined via
\be
\expval{a_1, a_2 | \rho_{A_1 A_2}^{T_{A_2}} | a_1', a_2'} = \expval{a_1, a_2' | \rho_{A_1 A_2}| a_1', a_2}
\ee
for basis states $\{ \ket{a_1} \}$ in $A_1$ and $\{ \ket{a_2} \}$ in $A_2$ \cite{VW, Plenio:2005cwa, Audenaert:2003}. The partial transpose is a positive but not completely positive map, which means some of the eigenvalues of $\rho_{A_1 A_2}^{T_{A_2}}$ (hereafter $\rho_{A_1 A_2}^{T_2}$) can be negative. Entanglement negativity quantifies the different between the eigenvalues of the partially transposed density matrix and the original density matrix via
\be
\mathcal{N}(\rho_{A_1 A_2} ) = \sum_{i} \frac{\abs{\lambda_i} - \lambda_i}{2} = \sum_{i: \lambda_i < 0} \abs{\lambda_i}.
\ee
As with the von Neumann entropy, there exist R\'enyi generalizations of entanglement negativity:
\be
\mathcal{N}_n = \Tr \left( \rho_{A_1 A_2}^{T_2} \right)^n.
\ee
Due to the absolute value, one needs to define two different analytic continuations for even and odd R\'enyi index $n$, so there are in fact two R\'enyi negativities given by
\ba
\mathcal{N}_{2k}^{\textrm{(even)}} &= \sum_{i} \abs{\lambda_i}^{2k} \nonumber \\
\mathcal{N}_{2k-1}^{\textrm{(odd)}}  &= \sum_i \textrm{sgn}{\lambda_i} \abs{\lambda_i}^{2k-1}
\ea
for integer $k$. We define relevant entanglement measures via analytic continuation from these quantities. The most common quantity to talk about is the logarithmic negativity, given via a $k \rightarrow 1/2$ analytic continuation of the even R\'enyi negativity
\be
\mathcal{E}(\rho_{A_1 A_2}) = \lim_{k \rightarrow 1/2} \log \mathcal{N}_{2k}^{\textrm{(even)}}(\rho_{A_1 A_2}) = \log \sum_i \abs{\lambda_i}.
\ee
One other quantity of interest is the partially transposed entropy, also known as the odd entropy, which is related to the $k \rightarrow 1$ analytic continuation of the odd R\'enyi negativity and is explicitly given by
\be
S^{T_2} \equiv \lim_{k \rightarrow 1} \frac{1}{2k - 2} \log \mathcal{N}_{2k-1} = - \sum_i \lambda_i \log \abs{\lambda_i}.
\ee
We need to include the R\'enyi entropy-like singular term out front as $\mathcal{N}^{\textrm{(odd)}}_1 = \Tr \rho_{A_1 A_2}^{T_2} = 1$.

\subsection{Disorder Averaged Negativity}
Now we can discuss the disorder average\footnote{In the condensed matter literature, disorder averaging has a different meaning and what we're doing should more properly be called ``ensemble averaging''. Ensemble averaging, however, already has a meaning in the high energy literature, so we keep with the terminology of \cite{Dong:2020iod}.} in the Gaussian approximation described in the previous section. The Schmidt decomposition of the energy eigenstate $\ket{E}$ is now
\be
\ket{E} = \sum_{i_1 j_1 J} M_{i j J}  \ket{E_{i}}_{A_1} \otimes \ket{E_{j}}_{A_2} \otimes \ket{E_J}_B.
\ee
Once again we'll consider $M_{i j J}$ as a Gaussian random variable, in particular with
\be
M_{i j J} \approx e^{-S(E)/2} C_{i j J}
\ee
\be
\overline{C_{i j J}} = 0, \quad \overline{C_{i j J} C_{i' j' J'}} = \delta_{i i'} \delta_{j j'} \delta_{J J'}.
\ee
The partially transposed density matrix is 
\be
\rho_{A_1 A_2}^{T_2} = \frac{1}{\mathcal{N}} \sum_{E_i E_j E_J} C_{i_1 j_1 J} C_{i_2 j_2 J} \ket{E_{i_1},E_{j_2} } \bra{E_{i_2} E_{j_1}},
\ee
or, by replacing dummy variables
\be
\rho_{A_1 A_2}^{T_2} = \frac{1}{\mathcal{N}} \sum_{E_i E_j E_J} C_{i_1 j_2 J} C_{i_2 j_1 J} \ket{E_{i_1},E_{j_1} } \bra{E_{i_2} E_{j_2}},  
\ee
where the sum over energies is understood to be in a window of width 3$\Delta$, though again we take this width to vanish. We also have
\be
\left( \rho_{A_1 A_2}^{T_2} \right)^n = \frac{1}{\mathcal{N}^n} \sum_{E_{i_1} , E_{j_1}}\sum_{i_1, \cdots, i_n, j_1, \cdots, j_n, J_1, \cdots J_n} \prod_{m=1}^{n} C_{i_m j_{m+1} J_m} C_{i_{m+1} j_m J_m} \ket{E_{i_1},E_{j_1} } \bra{E_{i_{n+1}} E_{j_{n+1}}}.
\ee
Note that the partial transpose has made it so the $i$ ($A_1$) indices are contracted cyclically, while the $j$ ($A_2$) indices are contracted anti-cyclically. The resolvent equation for these Wick contractions is the same as derived in \cite{Dong:2021oad}, and a more detailed explanation is given in Appendix \ref{app:resolvent}. We quote the result here:
\be
\lambda R(\lambda) = e^{S_{A_1} + S_{A_2}} + \frac{e^{S_B}}{e^{S_{A_2}}} \frac{R(\lambda)(1+R(\lambda))}{1-e^{2S_{A_2}}R(
\lambda)^2}.
\label{eq:resolvent}
\ee
\begin{figure}
    \centering
    \includegraphics[width=.6\textwidth]{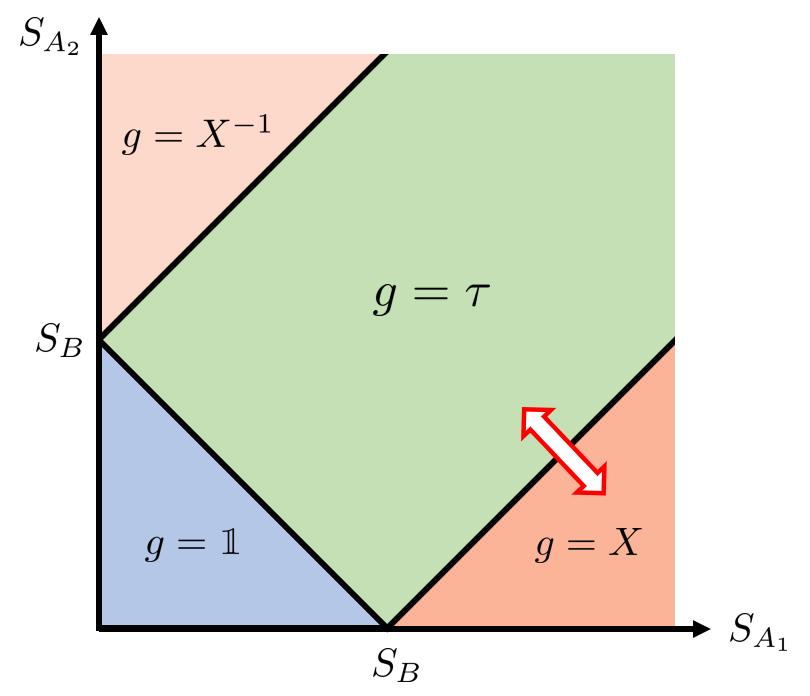}
    \caption{Phase diagram of R\'enyi negativity for various subsystem densities of state. The $g$'s label the dominant permutation which appears in the sum over Wick contractions; their exact forms are given in Appendix \ref{app:perm}. The resolvent equation \eqref{eq:resolvent} is valid in the regime $S_{A_2} << S_{A_1} + S_B$; we've indicated the forbidden region $g = X^{-1}$ in a lighter shade. Reproduced with minor alterations from \cite{Dong:2021oad}}
    \label{fig:phase}
\end{figure}
This resolvent equation furnishes a negativity spectrum described by the phase diagram in Figure \ref{fig:phase}. There are two transitions to consider. The first is when the $A$ and $B$ subsystems are the same size, i.e. $S_{A_1} + S_{A_2} = S_B$, corresponding to the transition from $g = \mathbbm{1}$ to $g = X$ in the phase diagram. From the calculation in \cite{Dong:2021oad}, we don't expect any enhanced corrections at this transition, so we don't study it in any detail, though the calculation would presumably follow the same steps. The second transition of interest is when the $A_1$ subsystem is the same size as the combined $A_2 B$ subsystem, $S_{A_1} = S_{A_2} + S_B$, corresponding to the transition from $g = \tau$ to $g = X$ in the phase diagram. In this regime the sum over diagrams is known explicitly, and the disorder averaged partially transposed density matrices are
\be
\Tr \overline{(\rho^{T_2}_{A_1 A_2})^{2k}} = \begin{cases}
\frac{1}{\mathcal{N}^{2k}} e^{2k(S_{A_2}+S_B) + S_{A_1}}e^{S_{A_2}}{}_2F_1 \left( 1-k, -2k; 2; e^{S_{A_1}-S_{A_2}-S_B} \right), S_{A_1} < S_{A_2} + S_B\\
\frac{1}{\mathcal{N}^{2k}} e^{2kS_{A_1} + S_{A_2}+S_B}e^{S_{A_2}}{}_2F_1 \left( 1-k, -2k; 2; e^{S_{A_2}+S_B-S_{A_1}} \right), S_{A_1} > S_{A_2} + S_B
\label{eq:evenTrrhoT}
\end{cases}
\ee
for even $n = 2k$ and 
\be
\Tr \overline{(\rho^{T_2}_{A_1 A_2})^{2k-1}} = \begin{cases}
\frac{1}{\mathcal{N}^{2k-1}} e^{(2k-1)(S_{A_2} + S_B) + S_{A_1}}{}_2F_1 \left( 1-2k, 1-k; 1; e^{S_{A_1}-S_{A_2}-S_B} \right),  S_{A_1} < S_{A_2} + S_B \\
\frac{1}{\mathcal{N}^{2k-1}} e^{(2k-1)S_{A_1} + S_{A_2}+S_B}{}_2F_1 \left( 1-2k, 1-k; 1; e^{S_{A_2}+S_B-S_{A_1}} \right), S_{A_1} > S_{A_2} + S_B
\label{eq:oddTrrhoT}
\end{cases}
\ee
for odd $n = 2k-1$. We give derivations for these formulae in Appendix \ref{app:perm}; the gist is that we sum over all permutations which lie on a geodesic between two dominant regions in phase space. The permutations on this geodesic can be enumerated, and the previous formulae are functions whose moments reproduce the combinatoric factors for these permutations.

\section{Negativity Phase Transitions}
\label{sec:trans}

We can use \eqref{eq:evenTrrhoT} and \eqref{eq:oddTrrhoT} to understand the difference between the microcanonical and canonical R\'enyi negativities in a chaotic eigenstate, using much the same techniques as were used in \cite{Dong:2020iod}. We denote by $f_{A_1}$ the volume fraction of $A_1$ such that the na\"ive phase transition happens at $f_{A_1} = 1/2$. We also denote the volume fraction of $A_2$ by $f_{A_2}$ and use $f_A = f_{A_1} + f_{A_2}$ to denote the total volume fraction of system $A$.

We impose energy conservation in all three subsystems, such that our ansatz is for subsystem entropies is
\ba
S_{A_1}(E_{A_1}) &= f_{A_1}V s \left( \frac{E_{A_1}}{f_{A_1}V}\right) \nonumber \\
S_{A_2}(E_{A_2}) &= f_{A_2}V s \left( \frac{E_{A_2}}{f_{A_2}V}\right) \nonumber \\
S_B(E - E_{A_1} - E_{A_2}) &= (1-f_A)V s \left( \frac{E - E_{A_1} - E_{A_2}}{(1-f_A)V}\right).
\label{eq:ansatz}
\ea
These again follow from ergodicity and imposing that the subsystem entropy is only a function of the subsystem energy density. 

\subsection{Even R\'enyi Negativity}

We'll start with studying the even R\'enyi negativities, from which the logarithmic negativity descends. Our expressions for the logarithms of the canonical ensemble and microcanonical ensemble R\'enyi negativities using our previous ansatzes are as follows:
\ba
\overline{\mathcal{N}_{2k}} &= \frac{1}{\mathcal{N}^{2k}} \int dE_{A_1} dE_{A_2}  e^{S_{A_1}(E_{A_1})+2S_{A_2}(E_{A_2})+S_B(E - E_{A_1} - E_{A_2})} G_k(f_{A_1},f_{A_2},E_{A_1}, E_{A_2}) \nonumber \\
\mathcal{N}_{2k}^{MC} &= \frac{1}{\mathcal{N}^{2k}} \int dE_{A_1} dE_{A_2}  e^{S_{A_1}(E_{A_1})+S_{A_2}(E_{A_2})+2k(S_{A_2}(E_{A_2})+S_B(E - E_{A_1} - E_{A_2}))},
\label{eq:evenNegs}
\ea
where the function $G_k(f_{A_1},f_{A_2},E_{A_1}, E_{A_2})$ is defined as the $k$-dependent part of \eqref{eq:evenTrrhoT}
\be
G_k(f_{A_1},f_{A_2},E_{A_1},E_{A_2}) = \begin{cases}
e^{(2k-1)(S_{A_2}+S_B)}{}_2F_1 \left( 1-k, -2k; 2; e^{S_{A_1}-S_{A_2}-S_B} \right), S_{A_1} < S_{A_2} + S_B \\
e^{(2k-1)S_{A_1}}{}_2F_1 \left( 1-k, -2k; 2; e^{S_{A_2}+S_B-S_{A_1}} \right), S_{A_1} > S_{A_2} + S_B,
\end{cases}
\ee
and $\mathcal{N}$ (with no other sub/superscripts) is an overall normalization given by
\be
\mathcal{N} = \int dE_{A_1} dE_{A_2} e^{S_{A_1}(E_{A_1})+S_{A_2}(E_{A_2})+S_B(E - E_{A_1} - E_{A_2})}.
\ee
Whenever unspecified, the subsystem entropies should now be understood to be valued at the subsystem energies, which we only omit for notational clarity. We write the difference between the logarithms of these quantities as
\be
\log \overline{ \mathcal{N}_{2k}} - \log \mathcal{N}^{MC}_{2k} \equiv \log \left( \frac{\int dE_{A_1} dE_{A_2} \exp( F_1(E_{A_1}, E_{A_2}))}{\int dE_{A_1} dE_{A_2}  \exp(F_2(E_{A_1}, E_{A_2}))} \right),
\ee
where the functions $F_1(E_{A_1}, E_{A_2})$ and $F_2(E_{A_1}, E_{A_2})$ are defined via the corresponding integrands in \eqref{eq:evenNegs}. The strategy will be to find saddle points for $F_1(E_{A_1}, E_{A_2})$ and $F_2(E_{A_1}, E_{A_2})$ and use the relative behavior of those saddle points to determine the scaling of the correction at transition.

We have two coupled saddle point equations for each both functions, which are given by
\ba
s' \left( \frac{E_{1}^{(1)}}{f_{A_1}V} \right) &= s' \left( \frac{E - E_{1}^{(1)} - E_{1}^{(2)}}{(1-f_A)V} \right) - \frac{\partial_{E_{A_1}} G_k(f_{A_1},f_{A_2},E_1^{(1)},E_1^{(2)})}{G_k(f_{A_1},f_{A_2},E_1^{(1)},E_1^{(2)})} \nonumber \\
2s' \left( \frac{E_{1}^{(2)}}{f_{A_2}V} \right) &= s' \left( \frac{E - E_{1}^{(1)} - E_{1}^{(2)}}{(1-f_A)V} \right) - \frac{\partial_{E_{A_2}} G_k(f_{A_1},f_{A_2},E_1^{(1)},E_1^{(2)})}{G_k(f_{A_1},f_{A_2},E_1^{(1)},E_1^{(2)})} \nonumber \\
s' \left( \frac{E_{2}^{(1)}}{f_{A_1}V} \right) &= 2k s' \left( \frac{E- E_{2}^{(1)} - E_{2}^{(2)}}{(1-f_A)V} \right) \nonumber \\
(2k+1)s' \left( \frac{E_{2}^{(2)}}{f_{A_2}V} \right) &= 2k s' \left( \frac{E - E_{2}^{(1)} - E_{2}^{(2)}}{(1-f_A)V} \right),
\label{eq:negsaddleseven}
\ea
where the pair $\mathcal{E}_1 = (E_{1}^{(1)}, E_{1}^{(2)})$ denotes a saddle point for $F_1(E_{A_1}, E_{A_2})$, while $\mathcal{E}_2 = (E_{2}^{(1)}, E_{2}^{(2)})$ denotes the saddle point for $F_2(E_{A_1}, E_{A_2})$. As $F_2(E_{A_1}, E_{A_2})$ is a strictly concave function, there is only one global maximum. $F_1(E_{A_1}, E_{A_2})$ on the other hand can have two maxima, as $G_k(f_{A_1},f_{A_2},E_{A_1}, E_{A_2})$ is strictly nonmonotonic.
\begin{figure}
    \centering
    \includegraphics[width=.8\textwidth]{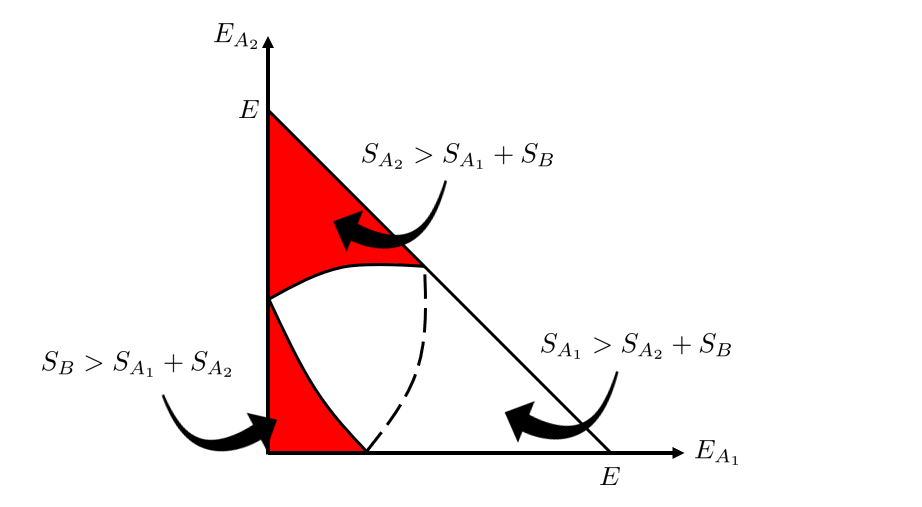}
    \caption{A schematic plot of regions (shaded in red) in the $E_{A_1} - E_{A_2}$ plane where our ansatz for the dominant sum over permutations does not hold. The lines separating the regions will depend sensitively on the form of  $s(e)$ and the volume fractions of the subsystems.}
    \label{fig:excluded}
\end{figure}

The first thing we have to be careful about is whether we are still within our regime of validity for probing the transition of interest. In the case of R\'enyi entropy, the fact that the dominant contribution comes from noncrossing partitions was assumed to hold for all of parameter space, that is for all values of subsystem entropy. This can be traced back to the fact that the dominant permutations all lie on a single geodesic $G(\mathbbm{1},X)$. In our case, we're trying to probe the transition on one geodesic $G(\tau,X)$ while suppressing diagrams from other geodesics, which imposes some natural constraints on the size of our subsystems.

We are justified in only considering the diagrams from Appendix \ref{app:perm} only if the saddle point energies satisfy the conditions:
\ba
S_{A_2}(E_{1,2}^{(2)}) < S_{A_1}(E_{1,2}^{(1)}) + S_B(E - E_{1,2}^{(1)} - E_{1,2}^{(2)}) \nonumber \\
S_B(E - E_{1,2}^{(1)} - E_{1,2}^{(2)}) < S_{A_1}(E_{1,2}^{(1)}) +  S_{A_2}(E_{1,2}^{(2)}),
\label{eq:phasespaceconds}
\ea
such that all contributions from subleading permutations remain subleading. We include a rough phase diagram of the allowed region to explore in Figure \ref{fig:excluded}. If the saddle point lies outside the allowed region, our answer for the dominant sum over permutations no longer holds, so we shouldn't try to explore those regions of phase space.

\begin{figure}
    \centering
    \includegraphics[width=.5\textwidth]{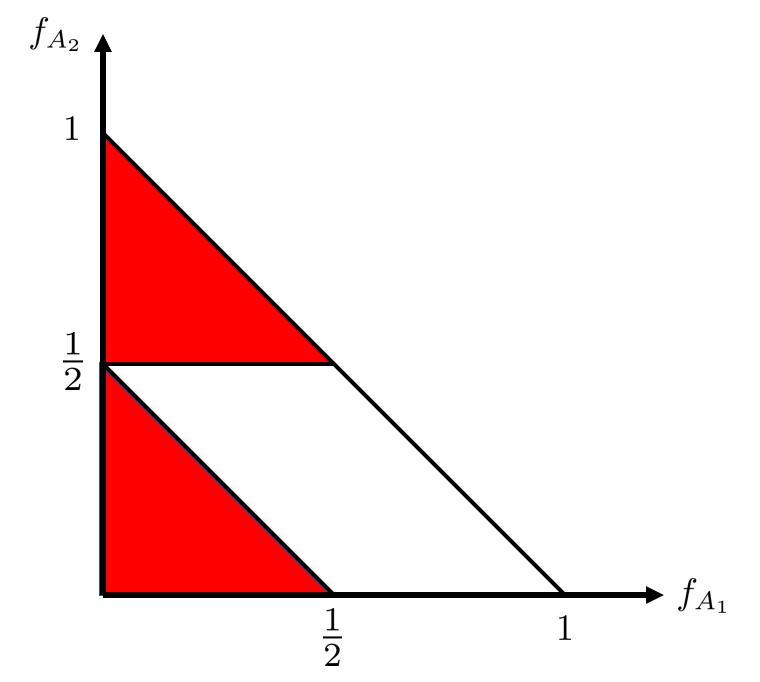}
    \caption{Excluded volume fractions from our analysis of the cyclic to pairwise phase transition. Describing the colored ``forbidden'' regions would require a sum over permutations we assert to be subdominant.}
    \label{fig:excludedf}
\end{figure}

This means before attempting to compute corrections at transition for all subsystem volume fractions, we should derive some bounds on the regime of validity of our approximation. We'll make use of the following inequality:
\be
S_{A_1}(E_{A_1}) + S_{A_2}(E_{A_2}) + S_B(E - E_{A_1} - E_{A_2}) \leq V s \left( \frac{E}{V} \right),
\ee
which follows from the fact that our subsystem entropy function $s(e)$ is concave. Plugging in the saddle points and using the first constraint in \eqref{eq:phasespaceconds} we can write
\ba
2 S_{A_2}(E_2^{(2)}) &< S_{A_1}(E_2^{(1)}) + S_{A_2}(E_2^{(2)}) + S_B(E -  E_2^{(1)} - E_2^{(2)}) < Vs \left( \frac{E}{V} \right) \nonumber \\
&\Rightarrow S_{A_2}(E_2^{(2)}) < \frac{V}{2}s \left( \frac{E}{V} \right).
\ea
We can use this relation to find
\ba
S_{A_2}(E_2^{(2)}) = f_{A_2} s \left( \frac{E_2^{(2)}}{f_{A_2}V}\right) > f_{A_2} s \left( \frac{E_2^{(2)}}{V} \right) \nonumber \\
\Rightarrow f_{A_2} s \left( \frac{E_2^{(2)}}{V} \right) < \frac{V}{2}s \left( \frac{E}{V}\right),
\ea
as $E_2^{(2)} < E$, a  constraint on $f_{A_2}$ which makes this true for all subsystem entropy densities is
\be
f_{A_2} < 1/2.
\ee
Therefore our calculations are only valid when subsystem $A_2$ is less than half of the total system size. We can find a similar inequality on $S_B$ using the second constraint in \eqref{eq:phasespaceconds}. We have
\ba
2S_B(E - E_2^{(1)} - E_2^{(2)}) &< S_{A_1}(E_2^{(1)}) + S_{A_2}(E_2^{(2)}) + S_B(E - E_2^{(1)} - E_2^{(2)}) \leq V s \left( \frac{E}{V} \right) \nonumber \\
\Rightarrow S_B(E - E_2^{(1)} - E_2^{(2)}) &< \frac{V}{2} s \left( \frac{E}{V} \right).
\ea
We can therefore write
\ba
S_B(E - E_2^{(1)} - E_2^{(2)}) = (1-f_A) V s \left( \frac{ E - E_2^{(1)} - E_2^{(2)} }{(1-f_A)V} \right) &> (1-f_A) V s \left( \frac{ E - E_2^{(1)} - E_2^{(2)} }{V} \right) \nonumber \\
\Rightarrow (1-f_A) V s \left( \frac{E - E_2^{(1)} - E_2^{(2)}}{V}\right) &< \frac{V}{2} s \left( \frac{E}{V} \right).
\ea
Again, a result that makes this inequality true for all saddle point energies is 
\be
f_A > 1/2.
\ee
This ties together a nice family of restrictions: both subsystems $A_2$ and $B$ have to have volume fraction less then half of the system. We illustrate these constraints in Figure \ref{fig:excludedf}. This makes some sense, as we want to probe transitions dominated by the behavior of $A_1$ relative to the rest of the system. 

Another way of seeing there should be a restricted regime for our procedure is as follows: entanglement negativity is agnostic as to which subsystem $A_1$ or $A_2$ one applies the partial transpose to. This would of course result in an averaged density matrix trace symmetric under exchange of $S_{A_1}$ and $S_{A_2}$, which our expressions \eqref{eq:evenTrrhoT} and \eqref{eq:oddTrrhoT} are not. However, by writing a resolvent equation valid only in a certain parameter regime, we can no longer comfortably integrate over all energies. This is an important point because the deviations from the featureless case can in principle be of order the system size, and so corrections are not necessarily perturbative as they were assumed to be in \cite{Dong:2021oad}.

We can, however, be comfortable in the validity of our calculation if the saddle points for $F_1(E_{A_1}, E_{A_2})$ and $F_2(E_{A_1}, E_{A_2})$ obey the conditions above, so restricting to the set of entropy functions which satisfy \eqref{eq:phasespaceconds}, let's first look at the saddle point equations for $\mathcal{E}_2$. Setting the third and fourth equations equal yields
\be
s' \left( \frac{E_2^{(1)}}{f_{A_1}V} \right) = (2k+1) s' \left( \frac{E_2^{(2)}}{f_{A_2}V} \right).
\ee
As $s'(e)$ is a monotonically decreasing function, for all $k > 0$ we have the inequality
\be
E_2^{(2)} > \frac{f_{A_2}}{f_{A_1}} E_2^{(1)}.
\label{eq:E21E22}
\ee
We can use this inequality to write a simple inequality on $E_2^{(1)}$ by rewriting the $E_2^{(1)}$ saddle point equation as 
\be
s' \left( \frac{E_2^{(1)}}{f_{A_1}V}\right) > 2k s' \left( \frac{E - \frac{f_A}{f_{A_1}} E_2^{(1)} }{(1-f_A) V}\right).
\ee
Now we have an inequality which depends on $k$, as we can write
\be
E_2^{(1)} < f_{A_1}E, \quad k \geq 1/2.
\ee
Note that this result is also valid for $k = 1/2$, as the relation \eqref{eq:E21E22} is a strict inequality which is never saturated for positive $k$. We can use a similar strategy to write an inequality for $E_2^{(2)}$. Rewriting the $E_2^{(2)}$ equation with \eqref{eq:E21E22} yields
\be
(2k + 1) s' \left( \frac{E_2^{(2)}}{f_{A_2}V}\right) < 2k s' \left( \frac{E - \frac{f_A}{f_{A_2}} E_2^{(2)}}{1 - f_A} \right).
\ee
The resulting inequality has a slightly different $k$ dependence:
\be
E_2^{(2)} > f_{A_2} E, \quad k > 0.
\ee
The last inequalities we can write are those for the saddle point values of $S_{A_1}$ and $S_{A_2}$:
\ba
S_{A_1}(E_2^{(1)}) &= f_{A_1}V s \left( \frac{E_2^{(1)}}{f_{A_1}V}\right) < f_{A_1}V s \left( \frac{E}{V}\right) \nonumber \\
S_{A_2}(E_2^{(2)}) &= f_{A_2}V s \left( \frac{E_2^{(2)}}{f_{A_2}V}\right) > f_{A_2} V s \left( \frac{E}{V}\right).
\label{eq:inequal1}
\ea
We'd like to find conditions on the hypergeometric being stuck on the first branch, i.e. $S_{A_1} < S_{A_2} + S_B$. This is guaranteed to happen if the weaker inequality $S_{A_1} < S_{A_2}$ is satisfied, which from \eqref{eq:inequal1} is necessarily true when
\be
f_{A_2} > f_{A_1}.
\ee
If we assume $S_{A_1} < S_{A_2}$ for the $\mathcal{E}_1$ saddle point as well, the argument of the hypergeometric is exponentially suppressed and we can approximate it by
\be
{}_2F_1(1-k,-2k;2;x) \approx 1 + k(k-1)x,
\ee
where the small parameter $x$ is now
\be
x \equiv e^{S_{A_1}(E_1^{(1)})-S_{A_2}(E_1^{(2)}) - S_B(E - E_1^{(1)} - E_1^{(2)})}.
\ee
Under this assumption the saddle point equations for $\mathcal{E}_1$ and $\mathcal{E}_2$ are the same up to exponentially suppressed terms, and therefore the saddle points $\mathcal{E}_1$ and $\mathcal{E}_2$ are exponentially close. This leads to the following form of corrections to ETH:
\be
\log \overline{\mathcal{N}_{2k}} - \log \mathcal{N}_{2k}^{MC} \propto \mathcal{O}(e^{-c V}), \quad k \geq 1/2, f_{A_2} > f_{A_1}
\ee
We can write a similar inequality for which $S_{A_1} < S_B$ is always satisfied. We recall the $E_2^{(1)}$ saddle point equation:
\be
s' \left( \frac{E_{2}^{(1)}}{f_{A_1}} \right) = 2k s' \left( \frac{E- E_{2}^{(1)} - E_{2}^{(2)}}{(1-f_A)V} \right).
\ee
At $k = 1/2$ there's clearly an equality between the arguments of the functions on the right and left, so for $k \geq 1/2$ we have the inequality
\be
\frac{E_2^{(1)}}{fV} \leq \frac{E - E_2^{(1)} - E_2^{(2)}}{(1-f_A)V}, \quad k \geq 1/2.
\label{eq:k12}
\ee
We'd like to satisfy the inequality $S_{A_1} < S_B$, or
\be
f_{A_1}V s \left( \frac{E_2^{(1)}}{fV}\right) < (1-f_A) V s \left( \frac{E - E_2^{(1)} - E_2^{(2)}}{(1-f_A)V}\right).
\ee
This is always satisfied if
\be
f_{A_1} < 1 - f_A.
\ee
So far we have two constraints which carve out a corner of the phase space for all $k \geq 1/2$. 

Now let's try to find a condition such that $S_{A_1} > S_{A_2} + S_B$. Using our previous ansatz this condition is written as
\be
f_{A_1} s \left( \frac{E_2^{(1)}}{f_{A_1}V}\right) > f_{A_2} s \left( \frac{E_2^{(2)}}{f_{A_2}V}\right) + (1-f_A) s \left( \frac{E - E_2^{(1)} - E_2^{(2)}}{(1-f_A)V}\right).
\ee
For all $k > 0$ we can use \eqref{eq:E21E22} to rewrite this as
\be
(f_{A_1} - f_{A_2}) s \left( \frac{E_2^{(2)}}{f_{A_2}V}\right) > (1-f_A) s \left( \frac{E - E_2^{(1)} -E_2^{(2)}}{(1-f_A)V}\right).
\ee
Using the $E_2^{(2)}$ saddle point equation, there exists for $k > 0$:
\be
\frac{E_2^{(2)}}{f_{A_2}} > \frac{E - E_2^{(1)} - E_2^{(2)}}{1-f_A}.
\ee
Therefore, $S_{A_1} > S_{A_2} + S_B$ is always satisfied if
\be
f_{A_1} - f_{A_2}  > 1 - f_A \Rightarrow f_{A_1} > 1/2, \quad k > 0.
\ee
For these volume fractions the corrections to the R\'enyi negativity are extensive in the system size, as $\mathcal{E}_1$ and $\mathcal{E}_2$ have no relation:
\be
\log \mathcal{N}_{2k} - \log \mathcal{N}_{2k} \propto \mathcal{O}(V), \quad k > 0, f_{A_1} > 1/2.
\ee
\begin{figure}
    \centering
    \includegraphics[width=.6\textwidth]{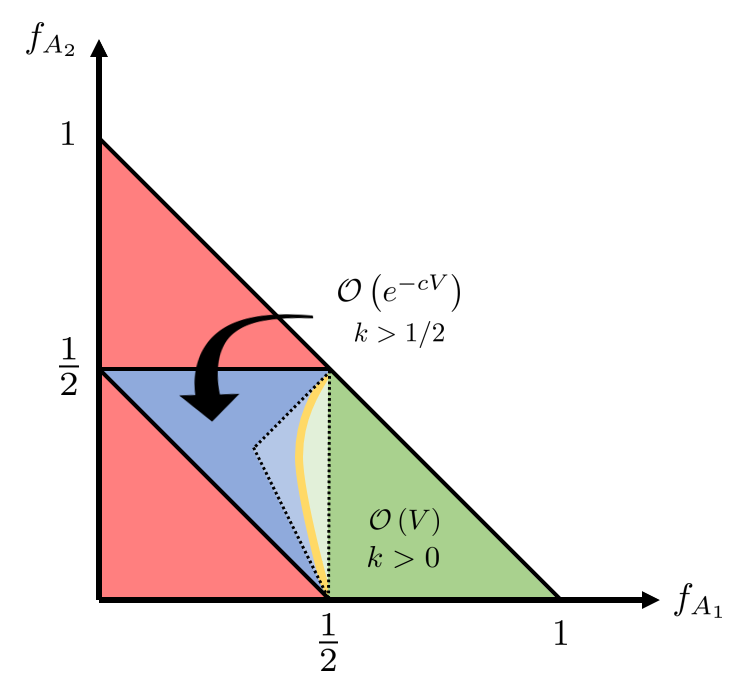}
    \caption{Phase diagram for corrections to even R\'enyi negativities. The concave region with $\mathcal{O}(e^{-cV})$ corrections comes from requiring $S_{A_1} < S_{A_2}$ and/or $S_{A_1} < S_B$. The $\mathcal{O}(V)$ region requires $S_{A_1} > S_{A_2} + S_B$. The interpolation between these regions will lie somewhere with $f_{A_1} < 1/2$ and is outlined by the dashed lines. The yellow curve represents a system specific boundary which will depend on $k$ and potentially on the specifics of $s(e)$.}
    \label{fig:evenphase}
\end{figure}We summarize the results so far in Figure \ref{fig:evenphase}. In that phase diagram, none of the boundaries should be though of as sharp, that is as \eqref{eq:E21E22} is never saturated for $k > 0$, neither are any constraints that depend on it. The interpolation between $\mathcal{O}(e^{-cV})$ corrections and $\mathcal{O}(V)$ corrections will happen somewhere in this ``unknown region'', though the only relevant point is that at $f_{A_1} = 1/2$ we should still be in a region with extensive corrections. In particular this implies the logarithmic negativity receives $\mathcal{O}(V)$ corrections, as was noted in \cite{Dong:2021oad}.\footnote{At $k = 1$, the even R\'enyi negativity is equal to the second R\'enyi entropy $S_2(\rho_A)$, which for $f_{A_1} + f_{A_2} > 1/2$ is expected to always receive volume law corrections, which we don't see for all volume fractions. This is an important consequence of the restriction to a particular phase transition; we require the full sum over diagrams to reproduce the partially transposed density matrix exactly.}

We won't comment on the case $k < 1/2$ for $f_{A_1} < 1/2$, though the expectation is that, like the $n < 1$ R\'enyi entropy, these measures always receive volume law corrections. It's also entirely possible the interpolating line continues moving towards the point $(0,1/2)$, meaning there's some set of volume fractions for which arbitrarily small but positive $k$ are well-approximated by ETH.

\subsection{Odd R\'enyi Negativity}
We can repeat the previous analysis for odd $n$. We have different expressions for the canonical and microcanonical R\'enyi negativities:
\ba
\log \overline{\mathcal{N}_{2k-1}} &= \frac{1}{\mathcal{N}_{2k-1}} \int dE_{A_1} dE_{A_2} e^{S_{A_1}(E_{A_1})+ S_{A_2}(E_{A_2})+S_B(E - E_{A_1} - E_{A_2})} G_k(f_{A_1}, f_{A_2}, E_{A_1}, E_{A_2}) \nonumber \\
\log \mathcal{N}_{2k-1}^{MC} &= \frac{1}{\mathcal{N}_{2k-1}} \int dE_{A_1} dE_{A_2} e^{S_{A_1}(E_{A_1})+ (2k-1)(S_{A_2}(E_{A_2})+S_B(E - E_{A_1} - E_{A_2}))},
\ea
where $G_k(f_{A_1},f_{A_2},E_{A_1}, E_{A_2})$ is now defined by \eqref{eq:oddTrrhoT} as:
\be
G_k(f_{A_1},f_{A_2},E_{A_1},E_{A_2}) = \begin{cases}
e^{(2k-2)(S_{A_2}+S_B)}{}_2F_1 \left( 1-k, 1-2k; 1; e^{S_{A_1}-S_{A_2}-S_B} \right), S_{A_1} < S_{A_2} + S_B \\
e^{(2k-2)S_{A_1}}{}_2F_1 \left( 1-2k, 1-k; 1; e^{S_{A_2}+S_B-S_{A_1}} \right), S_{A_1} > S_{A_2} + S_B.
\end{cases}
\ee
Again the subsystem entropies should be valued at their respective subsystem energies. Notably $\log \overline{\mathcal{N}_{2k-1}}$ enjoys a symmetry under $S_{A_1} \leftrightarrow S_{A_2} + S_B$. We again write the difference between the canonical and microcanonical answers as
\be
\log \overline{ \mathcal{N}_{2k-1}} - \log \mathcal{N}^{MC}_{2k-1} = \log \left( \frac{\int dE_{A_1} dE_{A_2} \exp( F_1(E_{A_1}, E_{A_2}))}{\int dE_{A_1} dE_{A_2}  \exp(F_2(E_{A_1}, E_{A_2}))} \right)
\ee
and use the same ansatz \eqref{eq:ansatz} to write the saddle point equations for $F_1$ and $F_2$ as
\ba
s' \left( \frac{E_{1}^{(1)}}{f_{A_1}V} \right) &= s' \left( \frac{E - E_{1}^{(1)} - E_{1}^{(2)}}{(1-f_A)V} \right) - \frac{\partial_{E_{A_1}} G_k(f_{A_1},f_{A_2},E_1^{(1)},E_1^{(2)})}{G_k(f_{A_1},f_{A_2},E_1^{(1)},E_1^{(2)})} \nonumber \\
s' \left( \frac{E_{1}^{(2)}}{f_{A_2}V} \right) &= s' \left( \frac{E - E_{1}^{(1)} - E_{1}^{(2)}}{(1-f_A)V} \right) - \frac{\partial_{E_{A_2}} G_k(f_{A_1},f_{A_2},E_1^{(1)},E_1^{(2)})}{G_k(f_{A_1},f_{A_2},E_1^{(1)},E_1^{(2)})} \nonumber \\
s' \left( \frac{E_{2}^{(1)}}{f_{A_1}V} \right) &= (2k-1) s' \left( \frac{E- E_{2}^{(1)} - E_{2}^{(2)}}{(1-f_A)V} \right) \nonumber \\
s' \left( \frac{E_{2}^{(2)}}{f_{A_2}V} \right) &= s' \left( \frac{E - E_{2}^{(1)} - E_{2}^{(2)}}{(1-f_A)V} \right).
\label{eq:negsaddlesodd}
\ea
Let's again investigate the saddle point for $F_2$. We immediately see
\be
\frac{E_2^{(2)}}{f_{A_2}} = \frac{E - E_2^{(1)} - E_2^{(2)}}{1-f_A}
\label{eq:A2equalsB}
\ee
for all $k$! This is a striking result, as it means we can write the sum of subsystem entropies in $A_2$ and $B$ as
\be
S_{A_2}(E_2^{(2)}) + S_B(E - E_2^{(1)}-E_2^{(2)}) \equiv S_{\overline{A_1}}(E_2^{(2)}) = (1-f_{A_1}) s \left( \frac{E_2^{(2)}}{f_{A_2}V} \right)
\ee
This is important as for the odd R\'enyi negativity, $S_{A_2}$ and $S_B$ always appear summed, so if we're only interested in the leading saddle point approximation we can treat them as one subsystem entropy $S_{\overline{A_1}}$. As such we can rewrite the single saddle point equation as
\be
s' \left( \frac{E_2^{(1)}}{f_{A_1}V}\right) = (2k-1) s' \left( \frac{E_2^{(2)}}{f_{A_2}V}\right).
\label{eq:F23}
\ee
We recognize this as similar to the saddle point equation \eqref{eq:renyisaddles} for $F_2(E)$ in the R\'enyi entropy, but we'll go through the discussion nonetheless. At $k = 1$ we can exactly solve for the subsystem energies and they are, unsurprisingly, proportional to the volume fractions of their respective subsystems:
\ba
E_2^{(1)} &= f_{A_1}E , \quad k = 1 \nonumber \\
E_2^{(2)} &= f_{A_2}E , \quad k = 1 .
\ea
When $k > 1$, we again have
\be
E_2^{(2)} > \frac{f_{A_2}}{f_{A_1}} E_2^{(1)},
\ee
which was true for general $k$ in the even case. Similar inequalities on volume fraction hold in the odd case; we still have
\ba
E_2^{(1)} &< f_{A_1}E , \quad k > 1 \nonumber \\
E_2^{(2)} &> f_{A_2}E , \quad k > 1 .
\ea
From this the inequality $S_{A_1} < S_{\overline{A_1}}$ is clearly satisfied when
\be
f_{A_1} < 1/2,
\ee
and corrections are exponentially suppressed. For $f_{A_1} > 1/2$, this won't be true generically and the corrections are extensive.

We can also say interesting things about $k < 1$. In this case the inequalities are flipped:
\ba
E_2^{(1)} &> f_{A_1}E , \quad k < 1 \nonumber \\
E_2^{(2)} &< f_{A_2}E , \quad k < 1 \nonumber \\
E_2^{(2)} &< \frac{f_{A_2}}{f_{A_1}} E_2^{(1)}.
\label{eq:dummy1}
\ea
We can check where $S_{A_1} > S_{\overline{A_1}}$. From \eqref{eq:dummy1} we have
\ba
S_{A_1}(E_2^{(1)}) &= f_{A_1}V s \left( \frac{E_2^{(1)}}{f_{A_1}V}\right) > f_{A_1}V s \left( \frac{E}{V}\right) \nonumber \\
S_{\overline{A_1}}(E_2^{(2)}) &= (1-f_{A_1})V s \left( \frac{E_2^{(2)}}{f_{A_2}V}\right) < (1-f_{A_1}) V s \left( \frac{E}{V}\right) .
\label{eq:inequal2}
\ea
We see that $S_{A_1} > S_{\overline{A_1}}$ is guaranteed to be satisfied if $f_{A_1} > 1/2$, and indeed there is no generic behavior for $f_{A_1} < 1/2$. Thus the corrections are extensive for all volume fractions for $k < 1$.

\subsection{Odd R\'enyi Negativity at Transition}

We would like to study this case in analogy with the entanglement entropy, for reasons that will be clear shortly. Let's follow the same procedure of dividing $F_1$ into two pieces, $F_{\textrm{dom}}$ and $F_\Delta$, defined as
\ba
F_{\textrm{dom}} &= S_{A_1} (E_{A_1}) + S_{A_2}(E_{A_2}) + S_B(E - E_{A_1} - E_{A_2}) \nonumber \\
&+ (2k-2)\textrm{max} \{ S_{A_1} (E_{A_1}), S_{A_2}(E_{A_2}) + S_B(E - E_{A_1} - E_{A_2}) \} \nonumber \\
F_\Delta &= \log {}_2F_1\left(1-2k,1-k;1;e^{-|S_{A_1} (E_{A_1})-S_{A_2}(E_{A_2}) - S_B(E - E_{A_1} - E_{A_2})|} \right).
\ea
That is, we take the dominant contribution and relegate the subleading contributions to a term bounded by $\mathcal{O}(1)$ in volume factors:
\be 
1 \leq e^{F_\Delta} \leq a_k, \quad a_{k} \equiv {3k-2 \choose k-1} = \frac{\Gamma(3k-1)}{\Gamma(k)\Gamma(2k)} = 1 + (k-1) + \mathcal{O}(k-1)^2.
\ee
The averaged R\'enyi negativity, with a $\frac{1}{2k-2}$ factor which will be important later, can be rewritten as
\be
\frac{1}{2k-2} \log \overline{\mathcal{N}_{2k-1}} = \frac{1}{2k-2} \log \left( \frac{1}{\mathcal{N}_{2k-1}} \int dE_{A_1} dE_{A_2} e^{F_{\textrm{dom}} + F_\Delta}  \right),
\ee
and we can bound $\log \overline{\mathcal{N}_{2k-1}}$ via
\be
\log \overline{\mathcal{N}_{2k-1}} - \log \mathcal{N}_{2k-1}^{\textrm{dom}} \leq \frac{1}{2} + \mathcal{O}(k-1).
\ee
As such $\mathcal{N}^{\textrm{dom}}_{2k-1}$ is enough to look for corrections larger than $\mathcal{O}(1)$.

Unlike the R\'enyi entropy, at $f = 1/2$ there's no obvious reflection symmetry of the energies in $F_{\textrm{dom}}$, and indeed we don't find one numerically. There is, however, a symmetry in the saddle points, which we'll argue for as follows. Call the two saddle points for $F_{\textrm{dom}}$ (or $F_1$, it makes no difference here) $\mathcal{E}_1^{(a)} = (E_1^{(1,a)}, E_1^{(2,a)})$ and $\mathcal{E}_1^{(b)} = (E_1^{(1,b)}, E_1^{(2,b)})$. Under the exchange $S_{A_1} \leftrightarrow S_{\overline{A}_1}$, the saddles are swapped due to the symmetry of the odd R\'enyi negativity. It's clear then at $f_{A_1} = 1/2$ there exists the equivalence
\ba
\frac{E_1^{(1,a)}}{f_{A_1}} &= \frac{E_1^{(2,b)}}{f_{A_2}} \nonumber \\
\frac{E_1^{(1,b)}}{f_{A_2}} &= \frac{E_1^{(2,a)}}{f_{A_2}}.
\ea
This means that the two saddle points contribute with equal magnitude, which contributes an $\mathcal{O}(1)$ factor to the difference between the canonical and microcanonical negativities:
\be
\frac{1}{2k-2} \left( \log \mathcal{N}^{\textrm{dom}}_{2k-1} - \log \mathcal{N}^{MC}_{2k-1} \right) = \frac{\log 2}{2-2k} \sim \mathcal{O}(1)
\ee

\begin{figure}
    \centering
    \includegraphics[width=.48\textwidth]{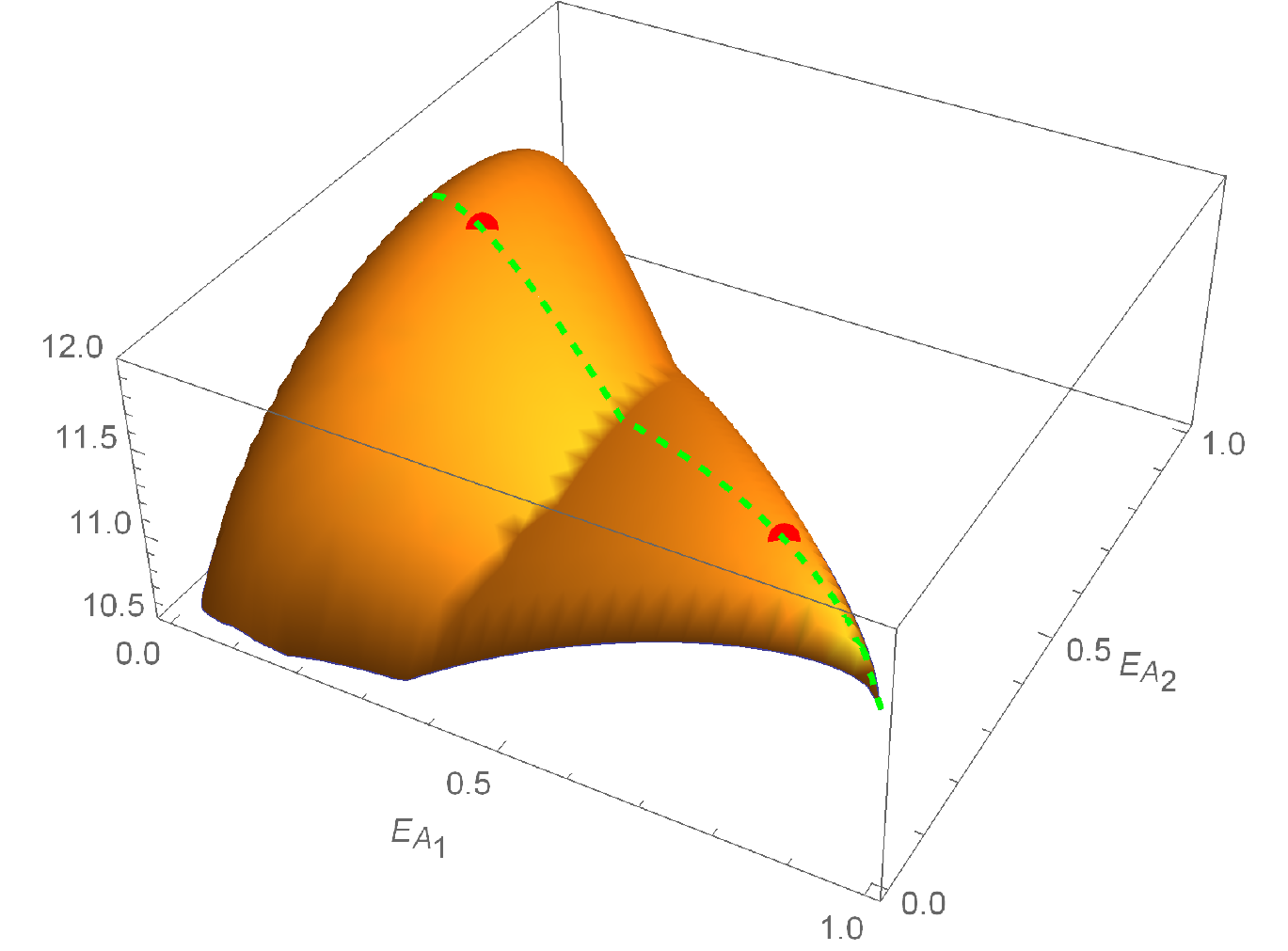}
    \includegraphics[width=.45\textwidth]{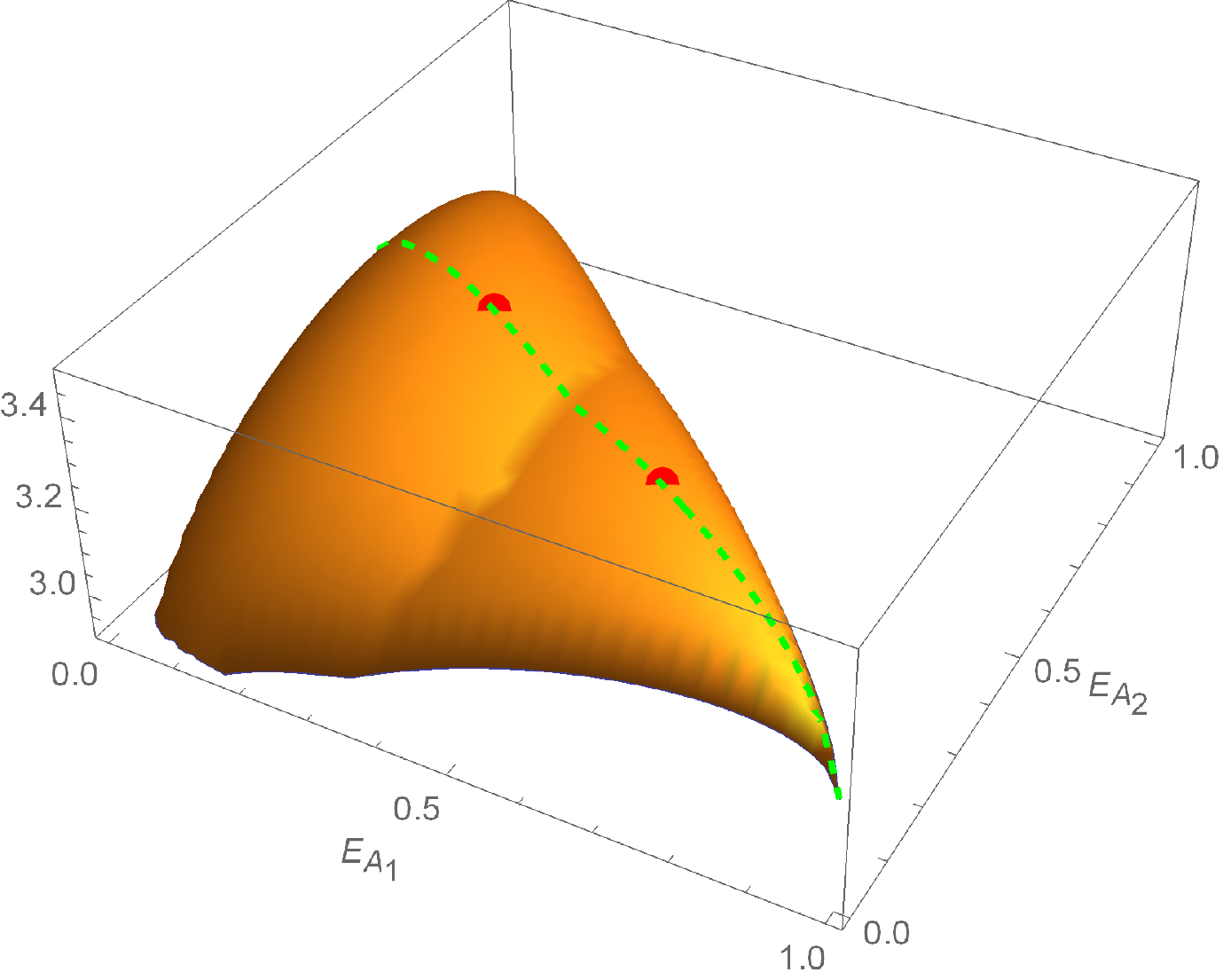}
    \includegraphics[width=.45\textwidth]{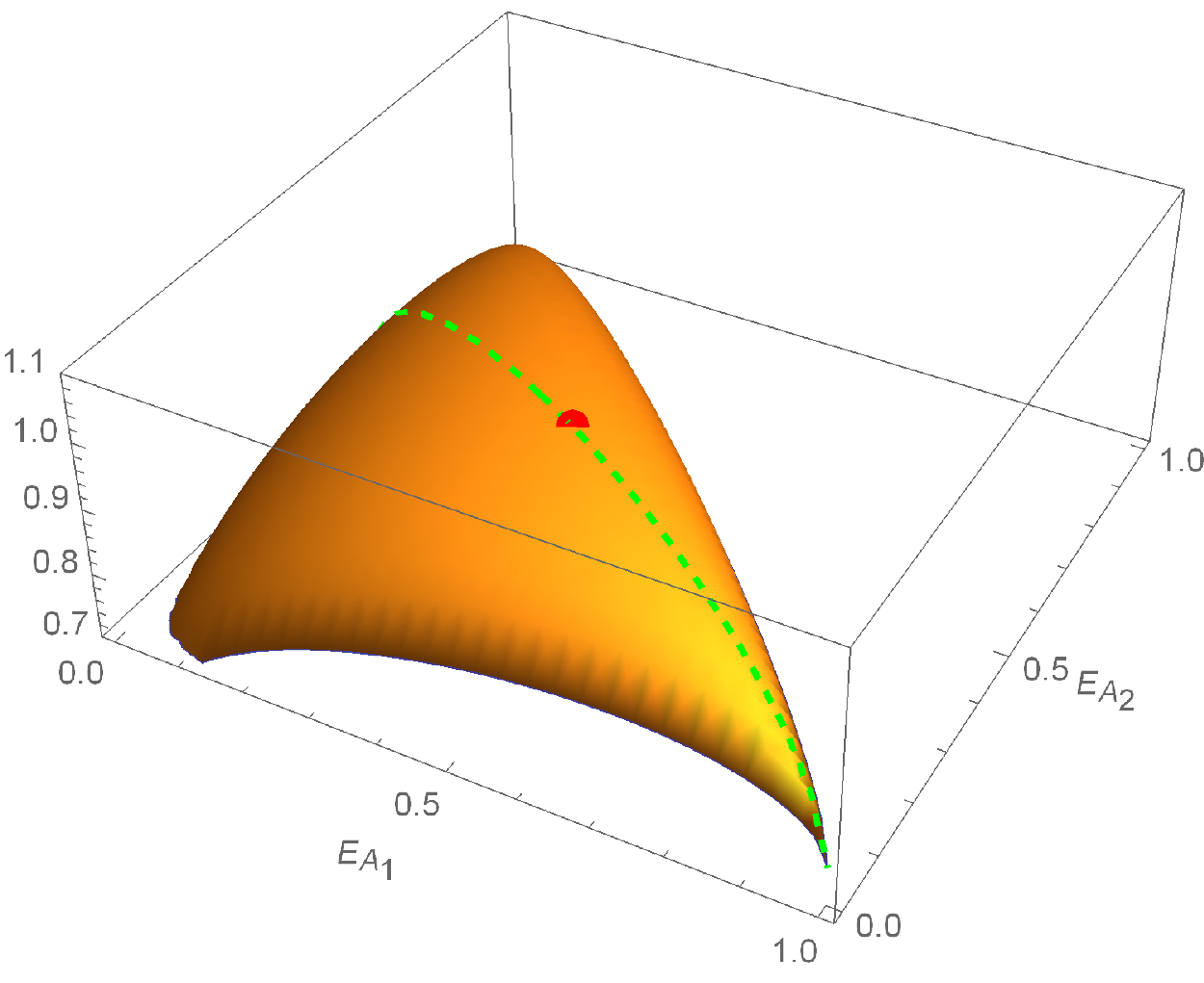}
    \caption{Plots of $F_1(E_{A_1}, E_{A_2})$ at phase transition. We've set $E = V = 1, f_{A_1} = 1/2$, and $f_{A_2} = 3/10$. For large $k$ (upper left), the two saddle points are well-separated  and can be treated separately. As we decrease $k$ (upper right) the saddle points approach one another and produce an emergent flat region. At exactly $k = 1$ (bottom) the saddle points coincide at $(f_{A_1}E, f_{A_2}E)$. The dotted line connecting the saddle points is given by $E_{A_2} = -2f_{A_2}(E_{A_1} - E)$; all saddles at $f_{A_1} = 1/2$ lie along this line.}
    \label{fig:odd}
\end{figure}

However, as in the case of von Neumann entropy, there is a subtletly related to the fact that the two saddles collide in the limit $k \rightarrow 1$, i.e. the partially transposed entropy. As they collide, there is an emergent region between the saddles which contributes to the integral, so we can't treat the presence of multiple equivalent saddles at leading order, we must integrate over the interpolating region. We show a plot of this phenomenon in Figure \ref{fig:odd}.
Let's solve the $F_2$ saddle point equations perturbatively in $\delta \equiv 2k-2$. The $E_2^{(1)}$ saddle point equation \eqref{eq:F23} becomes
\be
s' \left( \frac{E_2^{(1)}}{f_{A_1}V} \right) = (1+ \delta) s' \left( \frac{E_2^{(2)}}{f_{A_2}V}\right) \approx s' \left( \frac{E_2^{(2)}}{f_{A_2}V} +  \delta \frac{s' \left( E/V \right)}{s'' \left( E/V \right)} \right),
\ee
where we've again used that $E_2^{(2)} = f_{A_2}E$. Combining this with the unchanged \eqref{eq:A2equalsB} and plugging in $f_{A_1} = 1/2$ yields 
\ba
E_2^{(1)} &= \frac{E}{2} + \frac{V \delta}{4} \frac{s'(E/V)}{s''(E/V)} \nonumber \\
E_2^{(2)} &= f_{A_2}E - \frac{f_{A_2}V \delta}{2} \frac{s'(E/V)}{s''(E/V)}
\ea
From this we can write our subsystem entropies $S_{A_1}$ and $S_{\overline{A_1}}$ in the familiar form
\ba
S_{A_1}(E_2^{(1)}) &= \frac{1}{2} s \left( E + \frac{V \delta}{2} \frac{s'(E/V)}{s''(E/V)} \right) \nonumber \\
S_{\overline{A_1}}(E_2^{(2)}) &= \frac{1}{2} s \left( E - \frac{V \delta}{2} \frac{s'(E/V)}{s''(E/V)} \right)
\ea
What happens as $k \rightarrow 1$ for the odd R\'enyi negativity is precisely the same as what happens for the $n \rightarrow 1$ von Neumann entropy, namely that the $F_\Delta$ term ``fills in'' the space between the two saddles. The only difference is that this flat direction runs between two saddles separated along a line in the $E_{A_1} - E_{A_2}$ plane specified by $f_{A_2}$. The rest of the calculation is completely unchanged from that of the von Neumann entropy, and there is an enhanced correction exactly of the same form: 
\be
\overline{S^{T_2}} - S^{T_2}_{MC} = -\sqrt{\frac{C_V}{2 \pi}} + \mathcal{O}(\delta) \sim \mathcal{O}(\sqrt{V})
\ee
In \cite{Dong:2021oad}, it was noted that a na\"ive calculation shows the partially transposed entropy receives $\mathcal{O}(\sqrt{V})$ corrections, but a more accurate analysis shows it receives $\mathcal{O}(V)$ corrections. It would be interesting to understand the difference between our calculation and theirs.\footnote{A possible resolution is that our calculation was done at fixed $f_{A_2}$, roughly the same as fixing $k_2$ in \cite{Dong:2021oad}. Only when $k = k_1 k_2$ was fixed, similar to fixing $f_A$, do they see $\mathcal{O}(V)$ corrections.}

\section{Discussion and Future Work}

In this work we've studied a class of tripartite entanglement measures, the R\'enyi negativities, in a toy model of a chaotic eigenstate. We've resummed the relevant noncrossing permutations obtained via Wick contractions relevant at the transition of interest and studied the corrections to the dominant microcanonical saddle.

The main takeaway is as follows: logarithmic negativity and its R\'enyi generalizations thereof are not always ``good'' chaotic observables in the sense that their fluctuations (the difference between the canonical and microcanincal expectation values) are often of the same order as the quantities themselves, implying they are not self-averaging for all volume fractions. We've shown this is the case for the even R\'enyi negativity at transition, as well as for both even and odd R\'enyi negativities for $f_{A_1} > 1/2$. In particular we've shown that odd R\'enyi negativity behaves mostly the same as R\'enyi entropy at the $\tau$ to $X$ transition, exhibiting a $\mathcal{O}(\sqrt{V})$ enhanced correction at exactly $k = 1$. One surprising outcome is that, for both R\'enyi negativities, canonical typicality holds in some cases where the partially transposed density matrix is defined on a subsystem $A_1 A_2$ larger than half of the total system.

One interesting question is what bearing these volumetric corrections have on the validity of the cosmic brane prescription. It's expected that the holographic dual of subregion entanglement measures is given by the action of a geometric solution with a massive cosmic brane (or branes) inserted \cite{Lewkowycz:2013nqa, Dong:2016fnf}. Away from transition, it's expected that there is a single dominant saddle, or at the very least an $\mathcal{O}(1)$ number of equivalent saddles, all of which have have small enough fluctuations that we can treat the calculation of the brane area perturbatively. What happens if this saddle doesn't exist?\footnote{We thank Pratik Rath for discussions on this point.} For the $n < 1$ R\'enyi entropy, for example, the dual gravitational description is expected to be a cosmic brane with negative tension \cite{Dong:2016fnf}, so the minimal energy configuration would be a brane that falls towards the boundary. This is roughly the ``holographic dual'' of the $\mathcal{O}(V)$ corrections to ETH; it represents a failure of a single approximately geometric state to describe the dual system.

We now discuss some extensions to our work. A necessary restriction in our analysis is only summing over a subset of all relevant permutations near a particular phase transition. It would be useful to find a closed form expression for the moments of a block transposed Wishart matrix without these assumptions, which would involve finding a closed form solution to the recursion relation in \cite{banica2013asymptotic}. This would be especially nice as we could probe the region $f_A < 1/2$, which is where one could expect ETH to hold as the partially transposed density matrix is defined on less than half of the total system.

A technical point in our analysis was the use of 2-Dyck paths and 2-Narayana numbers, as opposed to (1-)Dyck paths which appear in the calculation of entanglement entropy. It's possible some further generalization of Narayana numbers (as in e.g. \cite{Novelli}) will be relevant for calculating transitions in higher party entanglement measures in a similar model.

So far, we've only discussed R\'enyi negativity, but there exists a family of holographically inspired measures termed refined R\'enyi negativities, which are given by
\ba
S^{T_2(n)}(\rho_{A_1 A_2}) = -n^2 \partial_n \left( \frac{1}{n} \log \mathcal{N}^{\textrm{(odd/even)}}_n (\rho_{A_1 A_2})\right)
\ea
We have not touched on the structure of transitions in these measures, but they could presumably be treated in the same way we've presented. Of particular interest is the refined R\'enyi 2-negativity $S^{T_2(2)}$, the $n \rightarrow 2$ limit of the even refined R\'enyi entropy. This quantity is explicitly given by
\be
S^{T_2(2)} = - \lim_{m \rightarrow 1} m^2 \partial_m \left( \frac{1}{m} \log \mathcal{N}^{\textrm{(even)}}_{2m} \right) = -\sum_i \frac{\lambda_i^2}{\sum_j \lambda_j^2} \log \left( \frac{\lambda_i^2}{\sum_j \lambda_j^2 }\right)
\ee
which is the von Neumann entropy of the normalized density matrix $\left( \rho^{T_2}_{A_1 A_2}\right)^2$. Consequently, the expectation is that the corrections will be $\mathcal{O}(\sqrt{V})$, which is indeed what is seen in the gravitational setting. It would be nice to derive this relation from our formalism.

Additionally, this formalism could be applied to study the reflected entropy \cite{Dutta:2019gen} and its R\'enyi generalizations thereof \cite{Akers:2021pvd, Akers:2022max, Akers:2022zxr}. Reflected entropy has been studied in a similar gravitational system \cite{Akers:2022max} and was shown to have $\mathcal{O}(\sqrt{V})$ corrections at transition, as in the case of the von Neumann entropy, derived via a resolvent calculation. Presumably the relevant permutations could be enumerated and the corrections calculated as we've done in this work.

We only considered the case where energy is conserved in all three subsystems. The authors of \cite{Vardhan:2021mdy} consider some cases in a similar model where some subystems are fixed at infinite temperature, which would correspond to freezing the density of states in those subsystems; it would be interested to understand to what extent this changes our results.

\section{Acknowledgements}

We thank Xi Dong, David Grabovsky, Jesse Held, Adolfo Holguin, Jonah Kudler-Flam, Ion Nechita, Pratik Rath, Mark Srednicki, and Wayne Weng for useful discussions. The work of SAM was supported by the Air Force Office of Scientific Research under award number FA9550-19-1-0360, the National Science Foundation under Grant No. PHY-1820908, and funds from the University of California. The work of FI was supported by an NSF Graduate Research Fellowship under Grant No. 2139319 and funds from the University of California.

\bibliographystyle{JHEP}
\bibliography{bibliography}

\appendix

\section{Deriving the Relevant Sum Over Permutations}
\label{app:perm}
Let's recall a few facts about the permutation group. For an element $g \in S_n$, we denote the number of swaps from the identity permutation $\mathbbm{1} = (1)(2)\cdots(n)$ to $g$ by $\ell(g)$ and the number of distinct cycles in $g$ by $\chi(g)$. These quantities satisfy the relation
\be
\ell(g) + \chi(g) = n
\ee
The number of swaps between two permutations $\ell(g^{-1}h) \equiv d(g,h)$ introduces a natural distance measure between two permutations. In particular, there exists the traingle inequality
\be
d(g,g_1) + d(g_1,h)  \geq d(g,h)
\ee
A \textit{geodesic} between two permutations $G(g,h)$ is the set of $g_1's$ which saturate this inequality. 

The sum over permutations we're interested in takes the form \cite{Shapourian:2020mkc, Dong:2021oad, Vardhan:2021mdy}
\be
\texttt{SUM} = \sum_{g \in S_n} (e^{S_B})^m (e^{S_{A_1}})^p (e^{S_{A_2}})^q
\ee
where we've made the substitutions
\be
m = \chi(g), \quad p = \chi(g^{-1}X), \quad q = \chi(g^{-1}X^{-1})
\ee
Here $X$ is the cyclic permutation $(1 2 \cdots n)$ and $X^{-1}$ is the anti-cyclic permutation $(n \; n-1 \cdots 1)$. This is the sum relevant for calculating the moments of a block transposed Wishart matrix \cite{banica2013asymptotic}, i.e. the weighting of Wick contractions when averaging over a random density matrix with Gaussian correlations. In our work, we're interested in the permutations which live on the geodesic $G(\mathbbm{1},X)$ and the geodesic $G(X,X^{-1})$ but not necessarily on the geodesic $G(\mathbbm{1},X^{-1})$. These permutations satisfy the following three equations:
\ba
m + p &= n + 1 \nonumber \\
m + q &\leq n + 1 \nonumber \\
p + q &= n + f(n)
\ea
where the function $f(n) = 1$ if $n$ is odd and $2$ if $n$ is even. When the second inequality is saturated, we're talking about the set of noncrossing pairings $\tau$. There are $C_n$ of these permutations, where $C_n$ are the Catalan numbers
\be
C_n = \frac{1}{n+1} \binom{2n}{n}
\ee
An example of a $\tau$ permutation on an even number of elements is $(12)(34)\cdots(n-1 \; n)$. A noncrossing pairing on an odd number of elements will have a single cycle of length 1 and all other cycles of length 2. Permutations which live on $G(\mathbbm{1}, X)$ and $G(X,X^{-1})$ but not $G(\mathbbm{1}, X^{-1})$ are precisely those which live on the single geodesic $G(\tau, X)$, which is the phase transition we're interested in. How do we enumerate these permutations? From the two equalities, we have:
\ba
p = n - 1 - m
q = m + f(n) - 1
\ea
From this, we can derive an upper bound on $m$:
\be
m \leq \frac{n + 2 - f(n)}{2}
\label{eq:mbound}
\ee
So we've reduced the sum over all permutations to a sum over a single parameter $m = \chi(g)$. We now have
\be
\texttt{SUM} = \sum_{m = 1}^{\frac{n-f(n)+2}{2}} T'(n, m)\left(e^{S_{A_2}}\right)^{f(n)-1} \left( e^{S_{A_1}} 
\right)^{n + 1}\left( \frac{e^{S_B}e^{S_{A_2}}}{e^{S_{A_1}}}\right)^m 
\ee
for some counting function $T'(n,m)$ which denotes the multiplicity at every $\chi(g)$. What is this function? Let's consider it for both even and odd $n$. For even $n = 2k$, the sum is
\be
\texttt{EVEN SUM} = \sum_{m = 1}^{k} T_e(k, m) e^{S_{A_2}} \left( e^{S_{A_1}}\right)^{2k + 1} \left( \frac{e^{S_B}e^{S_{A_2}}}{e^{S_{A_1}}}\right)^m 
\label{eq:evensum}
\ee
This is a sum over permutations starting with the cyclic permutation $X$ at $m = 1$ and ending with the pairwise connected permutations $\tau$ at $m = k$. Each $m$ corresponds to a permutation with $m$ cycles of even length. In this case, the numbers $T_e(k,m)$ are equivalent to the number of 2-Dyck paths of order $k$ with $m$ peaks and are given by
\be
T_e(k,m) = \frac{1}{k} \binom{k}{m} \binom{2k}{m-1}
\label{eq:Dyck2}
\ee
The $T_e(k,m)$ that appear here are analagous to the Narayana numbers which appear in the sum over non-crossing permutations. They are sometimes referred to as 2-Narayana numbers and appeared in various contexts elsewhere \cite{Brezin:1977sv, Novelli, Cachazo:2022voc}. We therefore have
\ba
\texttt{EVEN SUM} &= e^{S_{A_2}} \left( e^{S_{A_1}}\right)^{2k + 1} \sum_{m = 1}^{k} T_e(k,m) \left(\frac{e^{S_B}e^{S_{A_2}}}{e^{S_{A_1}}}\right)^m  \nonumber \\
&= e^{2k S_{A_1}} e^{2 S_{A_2}} e^{S_B} {}_2F_1 \left( 1-k, -2k, 2; \frac{e^{S_B}e^{S_{A_2}}}{e^{S_{A_1}}}\right)
\label{eq:even2F1}
\ea
Now let's look at the odd case. When $n = 2k - 1$, the sum over permutations is 
\be
\texttt{ODD SUM} = \sum_{m = 1}^{k} T_o(k, m) \left( e^{S_{A_1}} \right)^{2k} \left(\frac{e^{S_B}e^{S_{A_2}}}{e^{S_{A_1}}}\right)^m
\ee
Now the counting function is slightly different. We can derive it as follows: consider a permutation allowed in the even sum \eqref{eq:evensum} with $m$ cycles. The second binomial factor in \eqref{eq:Dyck2} can morally be thought of as choosing $m - 1$ distinct elements to belong to different cycles, while the rest is a symmetry factor that controls the number of noncrossing permutations modulo that choice. Therefore, one can think of each noncrossing permutation as living in a ``labelled'' superselection sector of size $\binom{2k}{m-1}$. By ignoring this choice and dividing by this factor, we can find a degenerate set of ``unlabelled'' noncrossing permutations. From this set we can remove an element from each cycle, so one unlabelled permutation in the even sum generates $m$ distinct unlabelled permutations in the odd sum, which then have to be relabelled to give the correct counting. This strategy of unlabelling, removing an element, and relabelling gives us
\be
T_o(k,m) = m \frac{\binom{2k-1}{m-1} }{\binom{2k}{m-1} } T_e(k,m) = \binom{2k-1}{m-1} \binom{k-1}{m-1}
\ee
\begin{figure}
    \centering
    \includegraphics[width=\textwidth]{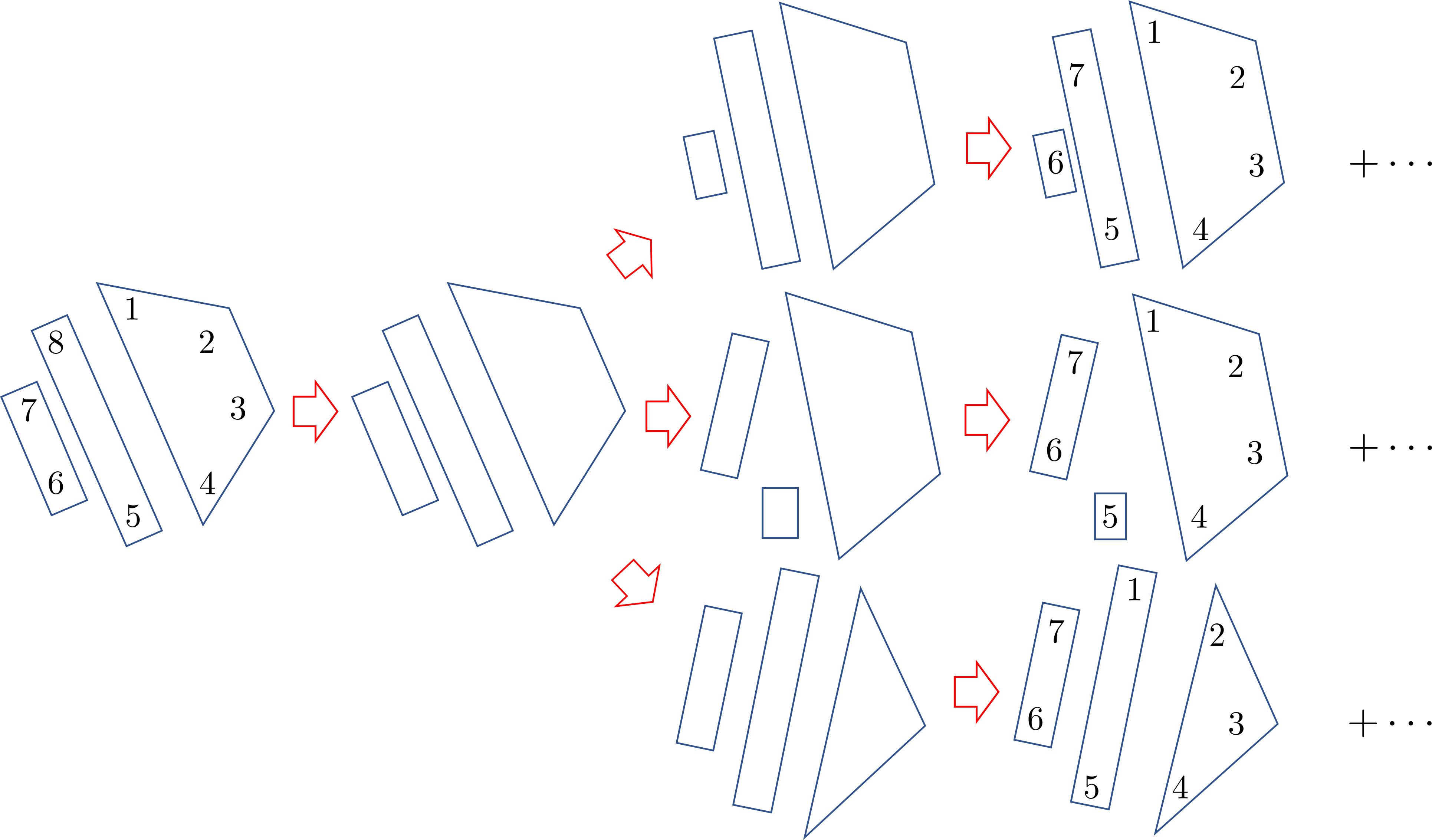}
    \caption{The procedure of generating permutations on $G(\tau,X)$ for odd $n$ from even $n$. We identify even permutations on $G(\tau,X)$ which are the same up to the choice of $m-1$ elements. Each of these pieces produces $m$ unlabelled odd pieces by removing an element, and identifying $m-1$ elements again gives us all odd permutations on $G(\tau,X)$.}
    \label{fig:perms}
\end{figure}If that was a bit too abstract, we illustrate this procedure in Figure \ref{fig:perms}. The sum over permutations for odd $n$ is now
\ba
\texttt{ODD SUM} &= \left( e^{S_{A_1}} \right)^{2k} \sum_{m = 1}^{k} T_e(k, m)  \left(\frac{e^{S_B}e^{S_{A_2}}}{e^{S_{A_1}}}\right)^m \nonumber \\
&= \left( e^{S_{A_1}} \right)^{2k-1} e^{S_{A_2}} e^{S_B} {}_2F_1\left( 1-2k, 1-k, 1; \frac{e^{S_B}e^{S_{A_2}}}{e^{S_{A_1}}} \right) 
\ea
As a sanity check for our counting functions $T_e(k,m)$ and $T_o(k,m)$, both $T_e(k,k)$ and $T_o(k,k)$ are equal to the symmetry factors for the pairwise connected geometries:
\be
T_e(k,k) = C_k, \quad T_o(k,k) = (2k-1)C_{k-1}
\ee
and $T_e(k,1) = T_o(k,1) = 1$.

\section{Resolvent for Disorder Averaging}
\label{app:resolvent}
The resolvent matrix $R_{ij}(\lambda)$ encodes the eigenvalue spectrum given by
\be
R(\lambda)_{ij} = \frac{1}{\lambda} \delta_{ij} + \sum_{n=1}^\infty \frac{1}{\lambda^{n+1}} \left( \rho_A^n \right)_{ij}
\ee
or in traced version
\be
R(\lambda) = \frac{e^{S_A}}{\lambda}  +  \sum_{n=1}^\infty \frac{1}{\lambda^{n+1}} \Tr \left( \rho_A^n \right)
\ee
There is a resolvent equation for the Wick contractions
\ba
R(\lambda)_{ij} = \frac{\delta_{ij}}{\lambda} + \frac{e^{S_B}}{\lambda} \sum_{n=1}^\infty R(\lambda)^n R(\lambda)_{ij} 
\ea
We can take the trace and resum this equation:
\be
\lambda R(\lambda) = e^{S_A} + \frac{e^{S_B} R(\lambda)}{1 - R(\lambda)}
\label{eq:resolv1}
\ee
This is the resolvent equation for the eigenvalues of a Wishart matrix. This equation can be solved exactly for $R(\lambda)$. It admits an expansion
\be
R(\lambda) = \frac{e^{S_A}}{\lambda} + \frac{e^{S_B}}{\lambda} \sum_{n=1}^\infty \sum_{k=1}^n N(n,k) \left( \frac{e^{S_A}}{\lambda}\right)^ne^{(k-1)(S_B-S_A)}
\ee
where $N(n,k)$ are the Narayana numbers. From this we can read off $\Tr \overline{ (\rho_A)^n}$:
\be
\Tr \overline{ (\rho_A)^n} = e^{nS_A + S_B} \sum_{k=1}^n N(n,k)e^{(k-1)(S_B - S_A)}
\ee
This sum has a nice closed form expression
\be
\Tr \overline{ (\rho_A)^n} = \begin{cases}
e^{S_A + nS_B} {}_2F_1 \left( 1-n, -n; 2; e^{S_A - S_B} \right), S_A < S_B \\
e^{nS_A + S_B} {}_2F_1 \left( 1-n, -n; 2; e^{S_B - S_A} \right), S_A > S_B
\end{cases}
\ee 
where there are two branches such that the final argument of the hypergeometric always always lies within the unit circle on the complex plane. A similar resolvent exists for the disorder averaging over Wick contractions for the partially transposed density matrix $\rho_{A_1 A_2}^{T_2}$. We work in the regime where $S_{A_2} \ll S_{A_1} + S_B$. The resolvent equation is \cite{Dong:2021oad}
\be
\lambda R(\lambda)_{j_1 j_2}^{i_1 i_2} = \delta^{i_1 i_2}_{j_1 j_2} + e^{S_B}\left( \sum_{m=1}^\infty 
\frac{R(\lambda)^{2m-2}}{e^{(2m-2)S_{A_2}}} R(\lambda)^{i_1 i_2}_{j_1 j_2} + \sum_{m=1}^\infty 
\frac{R(\lambda)^{2m-1}}{e^{(2m-2)S_{A_2}}} R(\lambda)^{i_1 i_2}_{j_1 j_2} \right)
\ee
Taking the trace:
\be
\lambda R(\lambda) = e^{S_{A_1} + S_{A_2}} + e^{S_B} \left( \sum_{m=1}^\infty \frac{R(\lambda)^{2m-1}(1+R(\lambda))}{e^{(2m-2)S_{A_2}}} \right)
\ee
and resumming gives the final resolvent equation
\be
\lambda R(\lambda) = e^{S_{A_1} + S_{A_2}} + \frac{e^{S_B}}{e^{S_{A_2}}} \frac{R(\lambda)(1+R(\lambda))}{1-e^{2S_{A_2}}R(
\lambda)^2}
\label{eq:cubicresolv}
\ee
We recognize this as the resolvent equation for the moments of a block transposed Wishart matrix. In the case $S_{A_2} = 0$ this reduces to the resolvent equation for the untransposed density matrix. This is a cubic equation and can be solved exactly, but the solution is not enlightening. \\

Previously, we saw that the moments of a Wishart matrix are given in closed form by a sum of Narayana numbers. An equivalent statement is that the inverse Stieltjes transform of the the resolvent defined by \eqref{eq:resolv1}, a generating function for the Narayana numbers, gives the eigenvalue spectrum of a Wishart matrix. The inverse Stieltjes transform of the solution to \eqref{eq:cubicresolv}, a generating function for the moments of a block transposed Wishart matrix (``block transposed Narayana numbers'') will produce the eigenvalue spectrum of a block transposed Wishart matrix (the ``negativity spectrum''). The block transposed Narayana numbers are not known in closed form; see \cite{banica2013asymptotic} for a recursive definition.

\end{document}